\documentclass{llncs}
\usepackage{amssymb} \usepackage{amsmath} \usepackage[all]{xy}
\usepackage{epsfig} \usepackage{color} \usepackage{epic}
\usepackage{eepic}

\newcommand{\var}{\operatorname{var}} 

\newcommand{\out}[1]{}

\newcommand{\uni}{~{\operatorname{UNION}}~}
\newcommand{\opt}{~{\operatorname{OPT}}~}

\newcommand{\andp}{~{\operatorname{AND}}~}

\newcommand{\vc}{~{\operatorname{FILTER}}~}

\newcommand{\AND}{\operatorname{AND}}
\newcommand{\OPT}{\operatorname{OPT}}
\newcommand{\UNION}{\operatorname{UNION}}
\newcommand{\FILTER}{\operatorname{FILTER}}

\newcommand{\bound}{\operatorname{bound}}
\newcommand{\isBlank}{\operatorname{isBlank}}
\newcommand{\isIRI}{\operatorname{isIRI}}

\newcommand{\dom}{\operatorname{dom}}
\newcommand{\M}{\Omega}
\newcommand{\mjoin}{\Join}
\newcommand{\mdif}{\smallsetminus}
\newcommand{\mpjoin}{\oj}

\newcommand{\Eval}{{\it Eval}}

\newcommand{\ql}{{\rm SPARQL}}

\newcommand{\sem}[1]{[\hspace*{-1pt}[#1]\hspace*{-1pt}]_D}
\newcommand{\semp}[2]{[\hspace*{-1pt}[#1]\hspace*{-1pt}]_{#2}}
\newcommand{\sproof}[1]{\noindent {\em Proof:}\ #1 \hfill$\Box$}
\newcommand{\aproof}[2]{\noindent {\em Proof of #1:}\ #2 \hfill$\Box$}
\newcommand{\eval}{\text{\sc Evaluation}}
\newcommand{\oj}{
\ \setlength{\unitlength}{0.00010000in}
\begin{picture}(924,639)(0,-10)
\path(312,612)(312,12)(912,612)
        (912,12)(312,612)(12,612)
\path(12,12)(312,12)
\end{picture}\ }
\newcommand{\aif}{{\tt if}}
\newcommand{\athen}{{\tt then}}
\newcommand{\areturn}{{\tt return}}

\begin{document}

\title{Semantics and Complexity of SPARQL}
\author{{\bf Jorge P\'erez}\inst{1} \and {\bf Marcelo Arenas}\inst{2}
\and {\bf Claudio Gutierrez}\inst{3}} 

\institute{Universidad de Talca \and
Pontificia Universidad Cat\'olica de Chile \and
Universidad de Chile}

\date{}
\maketitle

\begin{abstract}
SPARQL is the W3C candidate recommendation query language for RDF.
In this paper we address systematically the formal study
of SPARQL, concentrating in its graph pattern facility.
We consider for this study a fragment without literals and a simple
version of filters which encompasses all the main issues yet is
simple to formalize.
   We provide  a compositional semantics, prove there are
normal forms, prove complexity bounds, among others
that the evaluation of SPARQL patterns is PSPACE-complete,
compare our semantics to an alternative operational semantics,
give simple and natural conditions when both semantics coincide
and discuss optimizations procedures.
\end{abstract}

\section{Introduction}
   The Resource Description Framework (RDF)~\cite{RDFPrimer} 
is a data model for representing information about World Wide Web resources. 
Jointly with its release in 1998 as Recommendation of the W3C,
the natural problem of querying RDF data was raised.
Since then, several designs and implementations of RDF query languages
have been proposed  (see \cite{Haase} for a recent survey).
 In 2004 the RDF Data Access Working Group 
(part of the Semantic Web Activity) released a
first public working draft of a query language for RDF, 
called SPARQL~\cite{SPARQL}, whose specification
does not include RDF Schema.
  Currently (April 2006) SPARQL is a W3C Candidate Recommendation. 
 
  Essentially,  SPARQL is a graph-matching query language.
Given a data source $D$, a query consists of a pattern which is
matched against $D$, and the values obtained from this
matching are processed to give the answer.
 The data source $D$ to be queried can be composed of multiple sources.
    A SPARQL query consists of three parts.
  The {\em pattern matching part}, which includes 
several interesting features of pattern matching of graphs,
like optional parts, union of patterns, nesting,
filtering (or restricting) values of possible matchings,
and the possibility of choosing the data source to be matched
by a pattern.
 The  {\em solution modifiers}, which once 
the output of the pattern is ready
(in the form of a table of values of variables),  
allows to modify these values applying classical operators like
projection, distinct, order, limit, and offset.
 Finally, the {\em output} of a SPARQL query
can be  of different types:
yes/no queries, selections of values of the variables 
which  match the patterns,
construction of new triples from these values, and descriptions
 about resources queries. 

  Although taken one by one the features of SPARQL are simple
to describe and understand, it turns out that the combination of them
makes SPARQL into a complex language, whose semantics is far from
being understood. In fact,
the semantics of SPARQL currently given in the document~\cite{SPARQL},
as we show in this paper, does not cover all the
complexities brought by the constructs involved in SPARQL,
and includes  ambiguities, gaps and features difficult to understand. 
The interpretations of the examples and the semantics of cases not 
covered 
in \cite{SPARQL} are currently matter of long discussions 
in the W3C mailing lists. 

  The natural conclusion is that work on formalization of the
semantics of SPARQL is needed. 
  A formal approach to this subject is beneficial for several
reasons, including
to serve  as a tool to identify and derive relations among the
constructors, identify redundant and contradicting notions,
and to study the complexity, expressiveness, 
 and further natural database questions like rewriting and
optimization.
  To the best of our knowledge, there is no work today 
addressing this formalization systematically.  
 There are  proposals  addressing partial aspects of 
the semantics of some fragments of SPARQL.
   There are also  works addressing formal issues of the semantics
of query languages for RDF which can be of use for SPARQL. 
  In fact,  SPARQL shares several constructs with other proposals of 
query languages for RDF. In the related work section,
we discuss these developments in more detail.  
None of these works, nevertheless, covers the problems posed by
the core constructors of SPARQL from the syntactic, semantic,
algorithmic and computational complexity point of view, which is the
subject of this paper.

\paragraph{Contributions}

  An in depth analysis of the semantics benefits from abstracting 
some features, which although  relevant, in a first stage
tend to obscure the interplay of the basic constructors used in the
language. 
  One of our main goals was to isolate a core fragment of SPARQL
simple enough to be the subject matter of a formal analysis,
but which is expressive enough to capture the core complexities
of the language.  In this direction, we chose the
 graph pattern matching facility, which is
additionally one of the most complex parts of the language.
  The fragment isolated consists of the
grammar of patterns restricted to queries on one
dataset (i.e. not considering the dataset graph pattern)
over RDF without vocabulary of RDF Schema and literals.
  There are other two sources of abstractions which do not
alter in essential ways SPARQL: we use set semantics as opposed
to the bag semantics implied in the document of the W3C, and 
we avoid blanks in the syntax of patterns, because in our 
fragment can be replaced by variables~\cite{Mendel,Bruijn}.

The contributions of this paper are:

\begin{itemize}

\item A streamlined version of the core fragment of SPARQL with
       precise Syntax and Semantics.
 	A formal version of SPARQL helps clarifying cases where the
   current english-wording semantics gives little information, 
	identify areas of problems and permits to propose solutions.
       
\item We present a compositional semantics for patterns in SPARQL, 
prove that there
is a notion of normal form for graph patterns in the fragment considered,
 and indicate optimization procedures and rules for the operators
  based on them.

\item We give thorough analysis of the computational complexity of 
 the fragment. Among other bounds, we  prove that the complexity of 
evaluation of a general graph pattern in SPARQL
is PSPACE-complete even if we not consider filter conditions.

\item We formalize a natural procedural semantics which is implicitly
used by developers.  
 We  compare these two semantics, the operational and the compositional
mentioned above. We show that putting some slight and reasonable
syntactic restrictions on the scope of variables, they coincide, 
thus isolating a natural fragment 
having a clear semantics and an efficient evaluation procedure.



\end{itemize}

\subsection{Related Work}

\paragraph{Works on the \ql\ semantics.}
A rich source on the intended semantics of the constructors of SPARQL
are the discussions around W3C document~\cite{SPARQL}, which is still
in the stage of Candidate Recommendation. Nevertheless, systematic and
comprehensive approaches to define the semantics are not present, and
most of the discussion is based on use cases.  

   Cyganiak~\cite{hplabs1} presents a relational model of \ql.
  The author uses relational algebra operators (join, left outer join,
projection, selection, etc.) to model \ql\  \verb|SELECT| clauses.
  The central idea in~\cite{hplabs1} is to make a correspondence between
\ql\ queries and relational algebra queries over a single relation 
$T(S,P,O)$. Indeed a translation system between \ql\ and SQL is 
outlined. 
The system needs extensive use of \verb|COALESCE| and \verb|IS NULL| operations to resemble
\ql\ features.
  The relational algebra operators and their semantics
in~\cite{hplabs1} are similar to our operators and have similar 
syntactic and semantic issues.  
   With different motivations, but similar philosophy,
Harris~\cite{harris}  presents an implementation of \ql\  queries
in a relational database engine.
  He uses relational algebra operators similar to~\cite{hplabs1}.
 This line of work, which models the semantics of SPARQL 
based on the semantics of some relational operators, seems to be very
influent in the decisions on the W3C semantics of SPARQL.

  De Bruin et al.~\cite{Bruijn} address the definition of mapping
for SPARQL from a logical point of view. It slightly differs from
the definition in~\cite{SPARQL} on the issue of blank nodes.
 Although  De Bruin et al.'s definition allows blank nodes in graph patterns, 
it is similar to our definition which does not allow blanks in
patterns. In their approach, these blanks play
the role of ``non-distinguished'' variables, that is, variables
which are not presented in the answer.

  Franconi and Tessaris~\cite{franconi}, in an ongoing work on the
semantics of SPARQL,  formally define the solution for a basic graph pattern 
(an RDF graph with variables) as a set of partial functions.
 They also consider RDF datasets and several forms of RDF--entailment.
  Finally, they propose high level operators ($Join$, $Optional$, etc.) 
that take set of mappings and give set of mappings, 
but currently they do not have formal definitions for them,
stating only their types, i.e.,  the domain and codomain.

\paragraph{Works on semantics of RDF query languages.}

 There are several works on the semantics of RDF query languages which 
tangentially touch the issues addressed by SPARQL.
  Gutierrez et al.~\cite{Mendel} discuss the basic issues of the
semantics and complexity of a conjunctive query language for RDF with basic
patterns which underlies the basic evaluation approach of SPARQL.

 Haase et al.~\cite{Haase} present a comparison of functionalities 
of pre-SPARQL query languages, many of which served as inspiration
for the constructs of SPARQL. There is, nevertheless, no formal
semantics involved.

 The idea of having an algebraic query language for RDF is not new.
In fact, there are several proposals.
Chen et al.~\cite{Chen} present a set of operators for manipulating
RDF graphs, Frasincar et al.~\cite{RAL} study algebraic operators
on the lines of the RQL query language, and
Robertson~\cite{Triadic} introduces an algebra of triadic relations for RDF. 
Although they evidence the power of having an algebraic approach
to query RDF, the frameworks presented in each of these works makes
not evident how to model with them  the constructors of SPARQL.

 Finally, Serfiotis et al.~\cite{Griegos} study RDFS query fragments
using a logical framework, presenting results on the
classical database problems of  containment and minimization of queries for
a model of RDF/S. They concentrate on patterns using the RDF/S vocabulary
of classes and properties in conjunctive queries, making the overlap
with our fragment and approach almost empty.

\paragraph{Organization of the paper}
The rest of the paper is organized as follows.  Section 2 presents a
formalized algebraic syntax and a compositional semantics for SPARQL.
Section 3 presents the complexity study of the fragment considered.
Section 4 presents and in depth discussion of graph patterns not including
the $\operatorname{UNION}$ operator. Finally, Section
\ref{sec-conclusions} presents some conclusions.
Appendix~\ref{app:proofs} contains detailed proofs of all important results.


\section{Syntax and Semantics of SPARQL}
 
In this section, we give an algebraic formalization of the core
fragment of SPARQL over simple RDF, that is, RDF without RDFS
vocabulary and literal rules. This allows us to take a close look at
the core components of the language and identify some of its
fundamental properties (for details on RDF formalization see
\cite{Mendel}, or \cite{Draltan} for a complete reference including
RDFS vocabulary).

 Assume there are pairwise disjoint infinite sets $I$, $B$, and $L$
(IRIs,
Blank nodes, and RDF literals, respectively).  A
triple $(v_1, v_2, v_3) \in (I \cup B) \times I \times (I \cup B \cup
L)$ is called an {\em RDF triple}.  In this tuple, $v_1$ is 
the {\em subject}, $v_2$ the {\em predicate} and $v_3$ the {\em
object}.  We denote by $T$ the union $I\cup B \cup L$. 
Assume additionally the existence of an infinite set 
$V$ of variables disjoint from the above sets.

\begin{definition}
An {\em RDF graph}~\cite{RDFAbstract} is a set of RDF triples.  In our
context, we  refer to an RDF graph as an {\em RDF dataset}, or
simply a {\em dataset}.
\end{definition}



\subsection{Syntax of \ql\ graph pattern expressions}

 In order to avoid ambiguities in the parsing,
we present the syntax of \ql\ graph patterns 
in a more traditional algebraic way,
using the binary operators
$\operatorname{UNION}$, $\AND$ and
$\operatorname{OPT}$, and  $\operatorname{FILTER}$.
We fully parenthesize expressions and make explicit the left
associativity of \verb|OPTIONAL| and the precedence of AND over
\verb|OPTIONAL| implicit in~\cite{SPARQL}.

 A \ql\ graph pattern expression is defined recursively as follows:
\begin{enumerate}
\item A tuple from $(T \cup V) \times (I \cup V) \times (T \cup V)$ is
a graph pattern (a {\em triple pattern}).  

\item If $P_1$ and $P_2$ are graph patterns, then expressions
$(P_1 \andp P_2)$, $(P_1 \opt P_2)$, and $(P_1 \uni P_2)$ are graph
patterns. 
\item If $P$ is a graph pattern and $R$ is a {\em \ql\ built-in}
condition, then the expression $(P \vc R)$ is a graph pattern. 
\end{enumerate}
A {\em \ql\ built-in} condition
is constructed
using elements of the set $V\cup T$ and constants, logical connectives
($\neg$, $\wedge$, $\vee$), inequality symbols ($<$, $\leq$, $\geq$,
$>$), the equality symbol ($=$), unary predicates like $\bound$,
$\isBlank$, and $\isIRI$, plus other features
(see~\cite{SPARQL} for a complete list).

 In this paper, we restrict to the fragment of filters 
where the built-in condition is a boolean combination of terms
 constructed by using $=$ and $\bound$, that is:
\begin{enumerate} 
\item If $?X, ?Y \in V$ and $c \in I \cup L$, then $\bound(?X)$, $?X =
c$ and $?X = ?Y$ are built-in conditions.

\item If $R_1$ and $R_2$ are built-in conditions, then $(\neg R_1)$,
$(R_1 \vee R_2)$ and $(R_1 \wedge R_2)$ are built-in conditions.
\end{enumerate}
Additionally, we assume that for $(P \vc R)$ the condition $\var(R)
\subseteq \var(P)$ holds, where $\var(R)$ and $\var(P)$ are the sets
of variables occurring in $R$ and $P$, respectively. Variables in $R$
not occurring in $P$ bring issues that are not computationally
desirable.  Consider the example of a built in condition $R$ defined
as $?X=?Y$ for two variables not occurring in $P$. What should be the
result of evaluating $(P \vc R)$? We decide not to address this
discussion here.

%
%

\subsection{Semantics of \ql\ graph pattern expressions}
To define the semantics of \ql\ graph pattern expressions, we need to
introduce some terminology.  A {\em mapping} $\mu$ from $V$ to $T$ is
a partial function $\mu: V \rightarrow T$.  Abusing notation, for a
triple pattern $t$ we denote by $\mu(t)$ the triple obtained by
replacing the variables in $t$ according to $\mu$.  The domain of
$\mu$, $\dom(\mu)$, is the subset of $V$ where $\mu$ is defined.  Two
mappings $\mu_1$ and $\mu_2$ are {\em compatible} when for all
$x\in\dom(\mu_1)\cap\dom(\mu_2)$, it is the case that $\mu_1(x)=\mu_2(x)$, i.e. when $\mu_1\cup\mu_2$ is also a mapping.  Note that two mappings
with disjoint domains are always compatible, and that the empty
mapping (i.e. the mapping with empty domain) $\mu_{\emptyset}$ is
compatible with any other mapping.  Let $\M_1$ and $\M_2$ be sets of
mappings. We define the join of, the union of and the difference between
 $\M_1$ and $\M_2$ as:
\begin{eqnarray*}
\M_1\mjoin\M_2 &=& \{\mu_1\cup\mu_2\;|\;\mu_1\in\M_1,
          \mu_2\in\M_2\text{ are compatible mappings}\}, \\
\M_1 \cup \M_2 &=& \{\mu \;|\; \mu \in \M_1 \text{ or } \mu \in  \M_2\},\\
\M_1\mdif\M_2 &=& \{\mu\in\M_1\;|\;\text{ for all }\mu'\in\M_2\text{, } 
              \mu\text{ and }\mu'\text{ are not compatible}\}. 
\end{eqnarray*}
 Based on the previous operators, we define the left outer-join as:
\begin{eqnarray*}
\M_1\mpjoin \M_2 &=& (\M_1 \mjoin \M_2) \cup (\M_1 \mdif \M_2).
\end{eqnarray*}
We are ready to define the semantics of graph pattern expressions 
as a function $\sem{\,\cdot\,}$ which takes a pattern expression  
an returns a set of mappings.
We follow the approach in~\cite{Mendel}
defining the semantics as the set of mappings that matches the dataset
$D$. For simplicity, in this work we assume all datasets are already
lean, i.e. (for simple RDF graphs) this means they do not have 
redundancies, which as is proved in~\cite{Mendel}, 
ensures that the property that for all patterns and datasets,
if $D \equiv D'$ then $\sem{P}=\semp{P}{D'}$.
  This issue is not discussed in~\cite{SPARQL}.

\begin{definition}\label{def:evaluation}
Let $D$ be an RDF dataset over $T$, $t$ a triple pattern and $P_1,
P_2$ graph patterns. Then the {\em evaluation} of a graph pattern
over 
$D$, denoted by
$\sem{\,\cdot\,}$, is defined recursively as follows:
\begin{enumerate}
\item 
$\sem{t}=\{\mu\;|\;\dom(\mu)=\var(t)$ and $\mu(t)\in D\}$, where
$\var(t)$ is the set of variables occurring in $t$.

\item $\sem{(P_1 \andp P_2)} =
\sem{P_1}\mjoin\sem{P_2}$ .

\item $\sem{(P_1 \opt P_2)} =
\sem{P_1} \mpjoin \sem{P_2}$.

\item $\sem{(P_1 \uni P_2)} =
\sem{P_1}\cup\sem{P_2}$.  
\end{enumerate}

\end{definition}
The semantics of FILTER expressions goes as follows.
 Given a mapping $\mu$ and a built-in condition $R$, we
say that $\mu$ satisfies $R$, denoted by $\mu \models R$, if:
\begin{enumerate}
\item $R$ is $\bound(?X)$ and $?X \in \dom(\mu)$;

\item $R$ is $?X = c$, $?X \in \dom(\mu)$ and $\mu(?X) = c$;

\item $R$ is $?X = ?Y$, $?X \in \dom(\mu)$, $?Y \in \dom(\mu)$ and $\mu(?X) =
\mu(?Y)$;

\item $R$ is $(\neg R_1)$, $R_1$ is a built-in condition, and it
is not the case that $\mu \models R_1$;

\item $R$ is $(R_1 \vee R_2)$,  $R_1$ and $R_2$ are built-in
conditions, and $\mu \models R_1$ or $\mu \models R_2$;

\item $R$ is $(R_1 \wedge R_2)$, $R_1$ and $R_2$ are built-in
conditions, $\mu \models R_1$ and $\mu \models R_2$.
\end{enumerate}

\begin{definition}
Given an RDF dataset $D$ and a FILTER expression $(P \vc R)$, 
\begin{eqnarray*}
\sem{(P \vc R)} &=& \{ \mu \in \sem{P} \mid \mu \models R\}.
\end{eqnarray*}
\end{definition}

\begin{example}\label{ex:semantics}
Consider the RDF dataset $D$:
{\footnotesize
\begin{center}
\begin{tabular}{llllclllr}
$D=\{$ & ($B_1$, & name, & paul), &\ \ \ \ \ & ($B_1$, & phone, & 777-3426), \\
& ($B_2$, & name, & john), && ($B_2$, & email, & john@acd.edu), \\
&($B_3$, & name, & george), && ($B_3$, & webPage, & www.george.edu), \\
& ($B_4$, & name, & ringo), && ($B_4$, & email, & ringo@acd.edu), \\
& ($B_4$, & webPage, & www.starr.edu), && ($B_4$, & phone, & 888-4537), & $\}$
\end{tabular}
\end{center}}

\noindent
The following are graph pattern expressions and their evaluations over
$D$ according to the above semantics:
\begin{enumerate}
\item $P_1=((?A,$ email, $?E) \opt (?A,$ webPage, $?W))$. Then

{\footnotesize
\begin{flushleft}
\begin{tabular}{ccc}
$\sem{P_1}=$ & \hspace{.5cm} &
\begin{tabular}{l|c|c|c|} \cline{2-4}
& $?A$ & $?E$ & $?W$ \\ \cline{2-4}
$\mu_1:$ & $B_2$ & john@acd.edu & \\
$\mu_2:$ & $B_4$ & ringo@acd.edu & www.starr.edu \\ \cline{2-4}
\end{tabular}
\end{tabular}
\end{flushleft}
}

\item $P_2=(((?A,$ name, $?N) \opt (?A,$ email, $?E)) \opt (?A,$
webPage, $?W))$. Then 

{\footnotesize
\begin{flushleft}
\begin{tabular}{ccc}
$\sem{P_2}=$ & \hspace{.5cm} &
\begin{tabular}{l|c|c|c|c|} \cline{2-5}
&         $?A$ & $?N$ & $?E$ & $?W$ \\ \cline{2-5}
$\mu_1:$ & $B_1$ & paul & & \\
$\mu_2:$ & $B_2$ & john & john@acd.edu & \\
$\mu_3:$ & $B_3$ & george & & www.george.edu \\
$\mu_4:$ & $B_4$ & ringo & ringo@acd.edu & www.starr.edu \\ \cline{2-5}
\end{tabular}
\end{tabular}
\end{flushleft}
}

\item $P_3=((?A,$ name, $?N) \opt ((?A,$ email, $?E)  \opt (?A,$
webPage, $?W)))$. Then 

{\footnotesize
\begin{flushleft}
\begin{tabular}{ccc}
$\sem{P_3}= $ \hspace{.5cm} &
\begin{tabular}{l|c|c|c|c|} \cline{2-5}
&         $?A$ & $?N$ & $?E$ & $?W$ \\ \cline{2-5}
$\mu_1:$ & $B_1$ & paul & & \\
$\mu_2:$ & $B_2$ & john & john@acd.edu & \\
$\mu_3:$ & $B_3$ & george & & \\
$\mu_4:$ & $B_4$ & ringo & ringo@acd.edu & www.starr.edu \\ \cline{2-5}
\end{tabular}
\end{tabular}
\end{flushleft}
}

Note the difference between $\sem{P_2}$ and $\sem{P_3}$.
These two examples show that $\sem{((A \opt B) \opt C)}\not=\sem{(A
\opt (B \opt C))}$ in general.

\item $P_4=((?A,$ name, $?N) \andp ((?A,$ email, $?E) \uni (?A,$
webPage, $?W)))$. Then 

{\footnotesize
\begin{flushleft}
\begin{tabular}{ccc}
$\sem{P_4}= $ & \hspace{.5cm} &
\begin{tabular}{l|c|c|c|c|} \cline{2-5}
&         $?A$ & $?N$ & $?E$ & $?W$ \\ \cline{2-5}
$\mu_1:$ & $B_2$ & john & john@acd.edu & \\
$\mu_2:$ & $B_3$ & george & & www.george.edu \\
$\mu_3:$ & $B_4$ & ringo & ringo@acd.edu &  \\
$\mu_4:$ & $B_4$ & ringo & & www.starr.edu \\ \cline{2-5}
\end{tabular}
\end{tabular}
\end{flushleft}
}

\item \label{ex:filter} $P_5=(((?A,$ name, $?N) \opt (?A,$ phone, $?P)) \vc
?P=$777-3426$)$. Then 

{\footnotesize
\begin{flushleft}
\begin{tabular}{ccc}
$\sem{P_5}= $& \hspace{.5cm} &
\begin{tabular}{l|c|c|c|c|} \cline{2-4}
&         $?A$ & $?N$ & $?P$  \\ \cline{2-4}
$\mu_1:$ & $B_1$ & paul & 777-3426   \\ \cline{2-4}
\end{tabular}
\end{tabular}
\end{flushleft}
}


\end{enumerate}
\end{example}


\subsection{A simple normal form for graph patterns}
\label{sec-uf}
We say that two graph pattern expressions $P_1$ and $P_2$ are
{\em equivalent}, denoted by $P_1 \equiv P_2$, if $\sem{P_1} = \sem{P_2}$
for every RDF dataset $D$. 

\begin{proposition}\label{pro-uni}
Let $P_1$, $P_2$ and $P_3$ be graph pattern expressions
and $R$ a built-in condition.  Then: 
\begin{enumerate}
\item $\andp$ and $\uni$ are associative and commutative.

\item $(P_1 \andp (P_2 \uni P_3)) \equiv ((P_1 \andp P_2) \uni (P_1 \andp
P_3))$.

\item  $(P_1 \opt (P_2 \uni P_3)) \equiv ((P_1 \opt P_2) \uni (P_1
\opt P_3))$.

\item $((P_1 \uni P_2) \opt P_3) \equiv ((P_1 \opt P_3) \uni (P_2 \opt P_3))$.  

\item 
$((P_1 \uni P_2) \vc R) \equiv ((P_1 \vc R) \uni (P_2 \vc R)).$

\end{enumerate}
\end{proposition}
The application of the above equivalences permits to translate
any graph pattern into an equivalent one of the form:
\begin{eqnarray}\label{eq-uf-sp}
P_1\ \ \uni\ \ P_2\ \ \uni\ \ P_3\ \ \uni\ \ \cdots\ \ \uni\ \ P_n,
\end{eqnarray}
where each $P_i$ ($1 \leq i \leq n$) is a $\operatorname{UNION}$-free
expression.
In Section~\ref{sec:union-free}, we study UNION-free graph pattern
expressions. 


\section{Complexity of Evaluating Graph Pattern Expressions}
 
 A fundamental issue in any query language is the
complexity of query evaluation and, in particular, what is the
influence of each component of the language in this complexity.
In this section, we address these issues for graph pattern expressions.

As it is customary when studying the complexity of the evaluation
problem for a query language, we consider its associated decision problem.
We denote this problem by \eval\ and we define it as follows: 

\begin{center}
\begin{tabular}{lcl}
INPUT &:& An RDF dataset $D$, a graph pattern $P$ and a mapping
$\mu$.\\
QUESTION &:& Is $\mu \in \sem{P}$?
\end{tabular}
\end{center}
We start this study by considering the fragment consisting of graph
pattern expressions constructed by using only $\operatorname{AND}$
and $\operatorname{FILTER}$ operators. This simple fragment is
interesting as it does not use the two most complicated operators
in SPARQL, namely $\operatorname{UNION}$ and
$\operatorname{OPT}$. Given an RDF dataset $D$, a graph pattern 
$P$ in this fragment and a mapping $\mu$, it is possible to efficiently
check whether $\mu \in \sem{P}$ by using the following
algorithm. First, for each triple $t$ in $P$, verify whether $\mu(t)
\in D$. If this is not the case, then return {\em false}. Otherwise,
by using a bottom-up approach, verify whether the expression
generated by instantiating the variables in $P$ according to $\mu$
satisfies the FILTER conditions in $P$. If this is the case, then
return {\em true}, else return {\em false}. Thus, we conclude that:
\begin{theorem}\label{theo-af}
\eval\ can be solved in time $O(|P| \cdot |D|)$ for graph pattern
expressions constructed by using only $\operatorname{AND}$ and
$\operatorname{FILTER}$ operators. 
\end{theorem}
We continue this study by adding to the above fragment the
$\operatorname{UNION}$ operator. It is important to notice that the
inclusion of $\operatorname{UNION}$ in SPARQL is one of the most
controversial issues in the definition of this language. In fact, in
the W3C candidate recommendation for SPARQL \cite{SPARQL}, one can
read the following: {\em ``The working group decided on this design and
closed the disjunction issue without reaching consensus. The objection
was that adding UNION would complicate implementation and discourage
adoption''}. In the following theorem, we show that indeed the
inclusion of $\operatorname{UNION}$ operator makes the evaluation
problem for SPARQL considerably harder: 
\begin{theorem}\label{theo-afu}
\eval\ is NP-complete for graph pattern expressions constructed by
using only $\operatorname{AND}$, $\operatorname{FILTER}$ and
$\operatorname{UNION}$ operators.   
\end{theorem}
We conclude this study by adding to the above fragments the
$\operatorname{OPT}$ operator. This operator is probably the most
complicated in graph pattern expressions and, definitively, the most
difficult to define. The following theorem shows that the evaluation
problem becomes even harder if we include the $\operatorname{OPT}$
operator:
\begin{theorem}\label{theo-cc}
\eval\ is PSPACE-complete for graph pattern expressions. 
\end{theorem}
It is worth mentioning that in the proof of Theorem \ref{theo-cc}, we
actually show that \eval\ remains  PSPACE-complete if we consider
expressions without FILTER conditions, showing that the main source of
complexity in SPARQL comes from the combination of
$\operatorname{UNION}$ and $\operatorname{OPT}$ operators.

When verifying whether $\mu \in \sem{P}$, it is natural to assume that the
size of $P$ is considerably smaller that the size of $D$. This
assumption is very common when studying the complexity of a query
language. In fact, it is named data-complexity in the database
literature \cite{V82} and it is defined as the complexity of the
evaluation problem for a fixed query. More precisely, for the case of
SPARQL, given a graph pattern expression $P$, the evaluation problem
for $P$, denoted by $\eval(P)$, has as input an RDF dataset $D$ and a
mapping $\mu$, and the problem is to verify whether $\mu \in
\sem{P}$. From known results for the data-complexity of first-order
logic \cite{V82}, it is easy to deduce that:
\begin{theorem}\label{theo-dc}
$\eval(P)$ is in LOGSPACE for every graph pattern expression $P$.
\end{theorem}


\section{On the Semantics of UNION-free Pattern Expressions}
\label{sec:union-free}

The exact semantics of graph pattern expressions has been largely
discussed on the mailing list of the W3C. 
 There seems to be two main approaches proposed to compute
answers to a graph pattern expression $P$.
The first uses an operational semantics and
consists essentially in the execution of a depth-first
traversal of the parse tree of $P$ and the use of the intermediate results to
avoid some computations. 
This approach is the one followed by ARQ~\cite{ARQ}
(a language developed by HPLabs) in the cases we test,
and by the W3C when evaluating graph pattern expressions
containing nested optionals~\cite{S06}. For instance,
the computation of the mappings satisfying $(A
\opt (B \opt C))$ is done by first computing the mappings that match $A$, then
checking which of these mappings match $B$, and for those who match
$B$ checking whether they also match $C$~\cite{S06}.
 The second approach, compositional in spirit and the one we advocate here, 
extends classical conjunctive query evaluation~\cite{Mendel} and
is based on a bottom up evaluation of the parse tree, borrowing 
notions of  relational algebra evaluation~\cite{hplabs1,harris}
 plus some additional features.

As expected, there are queries for which both approaches do not
coincide (see Section \ref{sec-dif} for examples).  However, both
semantics coincide in most of the ``real-life'' examples.  For
instance, for all the queries in the W3C candidate recommendation for
SPARQL, both semantics coincide~\cite{SPARQL}.  Thus, a natural
question is what is the exact relationship between the two approaches
mentioned above and, in particular, whether there is a ``natural''
condition under which both approaches coincide. In this section, we
address these questions: 
Section~\ref{sec-greedy} formally introduces the depth-first approach,
discusses some issues concerning it, and presents queries for which the two
semantics do not coincide;
Section~\ref{sec-rel} identifies a natural and simple condition under 
which these two semantics are equivalent;
Section~\ref{sec-norm} defines a normal form and simple optimization procedures
for patterns satisfying the condition of Section~\ref{sec-rel}

Based on the results of Section~\ref{sec-uf}, we concentrate in the
critical fragment of UNION-free graph pattern expressions.

\subsection{A depth-first approach to evaluate graph pattern
expressions} 
\label{sec-greedy}
As we mentioned earlier, one alternative to evaluate graph pattern
expressions is based on a ``greedy'' approach that computes the
mappings satisfying a graph pattern expression $P$ by traversing the
parse tree of $P$ in a depth-first manner and using the intermediate
results to avoid some computations. This evaluation includes at each
stage three parameters: the dataset, the subtree pattern of $P$ to be
evaluated, and a set of mappings already collected. Formally, given
an RDF dataset $D$,
the evaluation of pattern $P$ with the set of mappings $\Omega$,
denoted by $\Eval_D(P, \Omega)$, is a recursive function defined as
follows:  

\begin{tabbing}
$Eva$\=$l_D$($P$: graph pattern expression, $\Omega$: set of mappings)\\
\>  \aif\ $\Omega=\emptyset$ \athen\ \areturn($\emptyset$)\\ 
\>  \aif\ $P$ is a triple pattern $t$ \athen\ \areturn($\Omega\mjoin
\sem{t}$)\\  
\> \aif\ $P=(P_1 \andp P_2)$ \athen\ \areturn\
$\Eval_D(P_2,\Eval_D(P_1,\Omega))$\\ 
\> \aif\ $P=(P_1 \opt P_2)$ \athen\ \areturn\ $\Eval_D(P_1,\Omega) \oj
\Eval_D(P_2,\Eval_D(P_1,\Omega))$\\
\> \aif\ $P=(P_1 \vc R)$ \athen\ \areturn\ $\{\mu \in \Eval_D(P_1,\Omega)
\;|\; \mu\models R\}$
\end{tabbing}

\noindent
Then, the evaluation of $P$ against a dataset $D$, which we
denote simply by $\Eval_D(P)$, is  $\Eval_D(P,\{\mu_\emptyset\})$,
where $\mu_\emptyset$ is the mapping with empty domain.

\begin{example}\label{ex-w3copt}
Assume that $P = (t_1 \opt (t_2 \opt t_3))$, where $t_1$, $t_2$ and
$t_3$ are triple patterns. To compute $\Eval_D(P)$,
we invoke function $\Eval_D(P, \{\mu_\emptyset\})$. This function in
turn invokes function $\Eval_D(t_1, \{\mu_\emptyset\})$, which returns
$\sem{t_1}$ since $t_1$ is a triple pattern and $\sem{t_1}
\mjoin \{\mu_\emptyset\} = \sem{t_1}$, and then it invokes
$\Eval_D((t_2 \opt t_3), \sem{t_1})$. As in the previous case,
$\Eval_D((t_2 \opt t_3), \sem{t_1})$ first invokes $\Eval_D(t_2,
\sem{t_1})$, which returns $\sem{t_1} \mjoin \sem{t_2}$
since $t_2$ is a triple pattern, 
and then it invokes $\Eval_D(t_3, \sem{t_1}
\mjoin \sem{t_2})$. 
Since $t_3$ is a triple pattern, the latter
invocation returns $\sem{t_1} \mjoin \sem{t_2} \mjoin
\sem{t_3}$. Thus, by the definition of $\Eval_D$ we have that
$\Eval_D((t_2 \opt t_3), \sem{t_1})$ returns $(\sem{t_1} \mjoin
\sem{t_2}) \oj (\sem{t_1} \mjoin \sem{t_2} \mjoin
\sem{t_3})$. Therefore, $\Eval_D(P)$ returns
 \begin{eqnarray*}
    \sem{t_1} \oj \big((\sem{t_1} \mjoin \sem{t_2}) \oj (\sem{t_1} \mjoin
   \sem{t_2} \mjoin \sem{t_3})\big).
 \end{eqnarray*}
Note that the previous result coincides with the evaluation algorithm
proposed by the W3C for graph pattern $(t_1 \opt (t_2 \opt t_3))$
\cite{S06}, as we first compute the mappings that match $t_1$, then we
check which of these mappings match $t_2$, and for those who match
$t_2$ we check whether they also match $t_3$. Also note that the result
of $\Eval_D(P)$ is not necessarily the same as $\sem{P}$ since
$\sem{(t_1 \opt (t_2 \opt t_3))} = \sem{t_1} \oj (\sem{t_2} \oj
\sem{t_3})$. 
In Example~\ref{ex:nestopt} we actually show a case 
where the two semantics do not coincide.
\end{example}

\paragraph{Some issues on the depth-first approach}
\label{sec-dif}

There are two relevant issues to consider when using the depth-first
approach to evaluate SPARQL queries.  First, this approach is not
compositional. For instance, the result of $\Eval_D(P)$ cannot in
general be used to obtain the result of $\Eval_D((P' \opt P))$, or even
the result of $\Eval_D((P' \andp P))$, as $\Eval_D(P)$ results from the
computation of $\Eval_D(P,\{\mu_\emptyset\})$ while $\Eval_D((P' \opt
P))$ results from the computation of $\Omega = \Eval_D(P',
\{\mu_\emptyset\})$ and $\Eval_D(P, \Omega)$.  This can become a
problem in cases of data integration where global answers are obtained
by combining the results from several data sources; or when storing
some pre--answered queries in order to obtain the results of more
complex queries by composition.  Second, under the depth-first
approach some natural properties of widely used operators do not hold,
which may confuse some users.  For example, it is not always
the case that $\Eval_D((P_1
\andp P_2)) = \Eval_D((P_2 \andp P_1))$, violating the commutativity of
the conjunction and making the result to depend on the order of the
query. 




\begin{example}\label{ex:nestopt}
Let $D$ be the RDF dataset shown in Example~\ref{ex:semantics} and
consider the pattern $P=((?X,\, \text{name},\, \text{paul}) \opt ((?Y,\,
\text{name},\,\text{george}) \opt (?X,\, \text{email},\, ?Z)))$.
Then $\sem{P}=\{\, \{?X\to B_1\}\, \}$, that is, $\sem{P}$ contains
only one mapping. On the other hand, following the recursive
definition of $\Eval_D$ we obtain that $\Eval_D(P)=\{\, \{?X\to B_1,
?Y\to B_3\}\, \}$, which is different from $\sem{P}$.
\end{example}


\begin{example}[Not commutativity of $\AND$]\label{ex:andnotcom}
Let $D$ be the RDF dataset in Example \ref{ex:semantics}, 
$P_1=((?X,\, \text{name},\, \text{paul}) \andp ((?Y,\,
\text{name},\,\text{george}) \opt (?X,\, \text{email},\, ?Z)))$ and 
$P_2=(((?Y,\, \text{name},\,\text{george}) \opt (?X,\,
\text{email},\, ?Z)) \andp (?X,\, \text{name},\, \text{paul}))$.
Then $\Eval_D(P_1)=\{\, \{?X\to B_1, ?Y\to B_3\}\, \}$ while
$\Eval_D(P_2)=\emptyset$. Using the compositional semantics, we obtain
$\sem{P_1}=\sem{P_2}=\emptyset$.

 Let us mention that ARQ~\cite{ARQ} gives the same non-commutative evaluation.

\end{example}


%




\subsection{A natural condition ensuring $\sem{P} = \Eval_D(P)$}
\label{sec-rel}
If for a pattern $P$ we have that $\sem{P} = \Eval_D(P)$ for every RDF
dataset $D$, then we have the best of both worlds for $P$ as the
compositional approach gives a formal semantics to $P$ while the
depth-first approach gives an efficient way of evaluating it. Thus, it
is desirable to identify natural syntactic conditions on $P$ ensuring
$\sem{P} = \Eval_D(P)$. In this section, we introduce one such
condition.

One of the most delicate issues in the definition of a semantics for
graph pattern expressions is the semantics of $\operatorname{OPT}$
operator.  
A careful examination of the conflicting examples
reveals a common pattern:
A graph pattern $P$ mentions
an expression $P' = (P_1 \opt P_2)$ and a variable $?X$ occurring both
in $P_2$ and outside $P'$ but not occurring in $P_1$. For instance, in
the graph pattern expression shown in Example \ref{ex:nestopt}:
\begin{eqnarray*}
P &=& ((?X,\, \text{name},\, \text{paul}) \opt ((?Y,\,
\text{name},\,\text{george}) \opt (?X,\, \text{email},\, ?Z))),
\end{eqnarray*}
the variable $?X$ occurs both in the optional part of the sub-pattern
$P'$ $=$ $((?Y$, $\text{name}$, $\text{george})$ $\opt$ $(?X$,
$\text{email}$, $?Z))$ and outside $P'$ in the triple $(?X$,
$\text{name}$, $\text{paul})$, but it is not mentioned in $(?Y$,
$\text{name}$, $\text{george})$.

What is unnatural about graph pattern $P$ is the fact that $(?X,\,
\text{email},\, ?Z)$ is giving optional information for  $(?X,\,
\text{name},\, \text{paul})$ but in $P$ appears as
giving optional information for $(?Y,\,\text{name},\,\text{george})$.
In general, graph pattern expressions having the condition mentioned
above are not natural. In fact, no queries in the W3C candidate
recommendation for SPARQL \cite{SPARQL} exhibit this condition.  This
motivates the following definition:

\begin{definition}
A graph pattern $P$ is {\em well designed} if for every occurrence 
of a sub-pattern $P' = (P_1 \opt P_2)$ of $P$ and for every variable
$?X$ occurring in $P$, the following condition holds:   
\begin{center}
if $?X$ occurs both in $P_2$ and outside $P' $, then it also occurs in $P_1$. 
\end{center}
\end{definition}
Graph pattern expressions that are not well designed are
shown in Examples \ref{ex:nestopt} and \ref{ex:andnotcom}.
For all these patterns, the two semantics differ.
 The next result shows a fundamental property of well-designed
graph pattern expressions, and is a
  welcome surprise as a very simple restriction on graph
patterns allows the users of SPARQL to alternatively use any of the
two semantics shown in this section:


\begin{theorem}\label{teo:eval}
Let $D$ be an RDF dataset and $P$ a well-designed graph pattern
expression. Then $\Eval_D(P)=\sem{P}$.
\end{theorem}

\subsection{Well-designed  patterns and normalization}
\label{sec-norm}
  Due to the evident similarity between certain operators of SPARQL
and relational algebra,  a natural question is whether the
classical results of normal forms and optimization in relational
algebra are applicable in the \ql\ context.  
  The answer is not straightforward, at least for the case of
optional patterns and its relational counterpoint, the left outer join.
  The classical results about outer join
query reordering and optimization by Galindo-Legaria and
Rosenthal~\cite{galindo1} are not directly applicable
in the \ql\ context because they assume constraints on the 
relational queries that are rarely found in \ql.  The first and more
problematic issue, is the assumption on predicates used for joining
(outer joining) relations to be \emph{null-rejecting}~\cite{galindo1}.
In \ql, those predicates are implicit in the variables that the graph
patterns share and by the definition of compatible mappings they are
never \emph{null-rejecting}.  In~\cite{galindo1} the queries are also
enforced not to contain Cartesian products, situation that occurs often
in \ql\ when joining graph patterns that do not share variables.  
Thus, specific techniques must be developed in the \ql\ context.  

 In what follows we show that the property of a pattern being
well designed has important consequences for the 
  study normalization and optimization for a fragment of 
\ql\ queries. We will restrict in this section to
graph patterns without $\FILTER$.

We start with equivalences that hold between
sub-patterns of well-designed graph patterns.

\begin{proposition}\label{prop:andopt}
Given a well-designed graph pattern $P$, if the left hand sides of the
following equations are sub-patterns of $P$, then:
\begin{eqnarray}
(P_1 \andp (P_2 \opt P_3)) \equiv ((P_1 \andp P_2) \opt P_3), \\
((P_1 \opt P_2) \opt P_3) \equiv ((P_1 \opt P_3) \opt P_2). \label{eq2}
\end{eqnarray}
Moreover, in both equivalences, if one replaces in $P$ the left hand side by
the right hand side, then the resulting pattern is still well designed.
\end{proposition}


\noindent     
From this proposition plus associativity and commutativity of 
$\AND$, it follows:

\begin{theorem} \label{teo:norm}
Every well-designed graph pattern $P$ is equivalent to a pattern in
the following {\em normal form}: 
\begin{eqnarray}
\label{form}
 (\cdots(t_1 \andp \cdots \andp t_k) \opt O_1) \opt O_2)\cdots )\opt O_n),
\end{eqnarray}
where each $t_i$ is a triple pattern,  $n \geq 0$ and each $O_j$ has
the same form $(\ref{form})$.
\end{theorem}

\noindent
The proof of the theorem 
is based on term rewriting techniques.
The next example shows the benefits of using the above normal
form.


%

\begin{example}
Consider dataset $D$ of Example~\ref{ex:semantics} and well-designed
pattern $P=(((?X$, $\text{name}, ?Y) \opt (?X, 
\text{email}, ?E)) \andp (?X, \text{phone}, \text{888-4537}))$. The
normalized form of $P$ is $P'=(((?X, \text{name}, ?Y) \andp (?X,
\text{phone}, \text{888-4537}))\opt$ $(?X, \text{email}, ?E))$. The
advantage of evaluating $P'$ over $P$ follows from a simple counting
of maps. 

\end{example}

\paragraph{Two examples of implicit use of the normal form.}
There are implementations (not ARQ\cite{ARQ}) that do not permit
nested optionals, and when evaluating a pattern they first evaluate
all patterns that are outside 
optionals and then \emph{extend} the results with the matchings of
patterns inside optionals.  That is, they are implicitly using the
normal form mentioned above.  In~\cite{hplabs1}, when evaluating a
graph pattern with relational algebra, a similar assumption is
made.  First the join of all triple patterns is evaluated,
and then the optional patterns are taken into account. Again, this is
an implicit use of the normal form.

\section{Conclusions}
\label{sec-conclusions}
The query language SPARQL is in the process of standardization, and in
this process the semantics of the language plays a key role.  A
formalization of a semantics will be beneficial on several grounds:
help identify relationships among the constructors that stay hidden in
the use cases, identify redundant and contradicting notions, study the
expressiveness and complexity of the language, help in optimization,
etc.

In this paper, we provided such a formal semantics for the graph
pattern matching facility, which is the core of SPARQL.  We isolated a
fragment which is rich enough to present the main issues and favor a
good formalization.  We presented a formal semantics, made
observations to the current syntax based on it, and proved several
properties of it.  We did a complexity analysis showing that unlimited
used of OPT could lead to high complexity, namely PSPACE.  We
presented an alternative formal procedural semantics which closely
resembles the one used by most developers. We proved that under simple
syntactic restrictions both semantics are equivalent, thus having the
advantages of a formal compositional semantics and the efficiency of a
procedural semantics.  Finally, we discussed optimization based on
relational algebra and show limitations based on features of
SPARQL. On these lines, we presented optimizations based on normal
forms.

Further work should concentrate on the extensions of these ideas to the
whole language and particularly to the extension --that even the current
specification of SPARQL lacks-- to RDF Schema.


\appendix

\section{Proofs and Intermediate Results}\label{app:proofs}
\subsection{Some technical results}

\begin{lemma}\label{lem-oor}
All the following equivalences hold:
\begin{enumerate}
\item[(1)] If $P$ is a graph pattern and $R_1$, $R_2$ are built-in
conditions such that $\var(R_1) \subseteq \var(P)$ and $\var(R_2)
\subseteq \var(P)$, then 
\begin{eqnarray*}
((P \vc R_1) \vc R_2) &\equiv& (P \vc (R_1 \wedge R_2)),\\
(P \vc (R_1 \vee R_2)) &\equiv& ((P \vc R_1) \uni (P \vc
R_2)).
\end{eqnarray*}

\item[(2)] If $P_1$ and $P_2$ are conjunctions of triple patterns and
$R$ is a built-in condition such that $\var(R) \subseteq \var(P_1)$,
then 
\begin{eqnarray*}
((P_1 \vc R) \andp P_2) & \equiv & ((P_1 \andp P_2) \vc R).
\end{eqnarray*}
\end{enumerate}
\end{lemma}

\sproof{(1.1) Let $D$ be an RDF database. Assume first that $\mu \in
\sem{((P \vc R_1) \vc R_2)}$. Then $\mu \in \sem{(P \vc R_1)}$ and
$\mu \models R_2$. Thus, $\mu \in \sem{P}$, $\mu \models R_1$ and $\mu
\models R_2$. Therefore, $\mu \models (R_1 \wedge R_2)$ and, hence, we
conclude that $\mu \in \sem{(P \vc (R_1 \wedge R_2))}$. Now assume
that $\mu \in \sem{(P \vc (R_1 \wedge R_2))}$. Then $\mu \in \sem{P}$
and $\mu \models (R_1 \wedge R_2)$. Thus, $\mu \in \sem{P}$, $\mu
\models R_1$ and $\mu \models R_2$. We conclude that $\mu \in \sem{(P \vc
R_1)}$ and, therefore, given that $\mu \models R_2$, we have that $\mu
\in \sem{((P \vc R_1) \vc R_2)}$. 

(1.2) Given an RDF database $D$, we have that:
\begin{eqnarray*}
\sem{(P \vc (R_1 \vee R_2))} &=& \{\mu \in \sem{P} \mid \mu \models
(R_1 \vee R_2)\}\\ 
&=& \{\mu \in \sem{P} \mid \mu \models R_1 \text{ or } \mu \models
R_2)\}\\ 
&=& \{\mu \in \sem{P} \mid \mu \models R_1 \} \cup  \{\mu \in \sem{P}
\mid \mu \models R_2)\}\\  
&=& \sem{(P \vc R_1)} \cup \sem{(P \vc R_2)}\\
&=& \sem{((P \vc R_1) \uni (P \vc R_2))}.
\end{eqnarray*}

(2) Let $D$ be an RDF database. Assume first that $\mu \in \sem{((P_1
\vc R) \andp P_2)}$. Then there exist $\mu_1 \in \sem{(P_1 \vc R)}$
and $\mu_2 \in \sem{P_2}$ such that $\mu_1$ and $\mu_2$ are compatible
and $\mu = \mu_1 \cup \mu_2$. Since $\mu_1  \in \sem{(P_1 \vc R)}$, we
have that $\mu_1 \in \sem{P}$ and $\mu_1 \models R$. Given that $P_1$
is a conjunction of triple patterns and $\var(R) \subseteq \var(P_1)$, we
have that $\mu_1(?X)$ is defined for every $?X \in \var(R)$. Thus, given
that $\mu_1 \models R$ and $\mu_1$ is contained in $\mu$, we conclude
that $\mu \models R$. Therefore, given that $\mu_1 \in \sem{P_1}$ and
$\mu_2 \in \sem{P_2}$, we have that $\mu = \mu_1 \cup \mu_2 \in
\sem{(P_1 \andp P_2)}$ and, hence, $\mu \in \sem{((P_1 \andp P_2) \vc
R)}$. Now assume that $\mu \in \sem{((P_1 \andp P_2) \vc R)}$. Then $\mu
\models R$ and $\mu \in \sem{(P_1 \andp P_2)}$ and, therefore, there
exist $\mu_1 \in \sem{P_1}$ and $\mu_2 \in \sem{P_2}$ such that $\mu_1$ and
$\mu_2$ are compatible and $\mu = \mu_1 \cup \mu_2$.  Given that $(P_1
\andp P_2)$ is a conjunction of triple patterns and $\var(R) \subseteq
\var(P_1) \subseteq \var((P_1 \andp P_2))$, we have that $\mu(?X)$ is
defined for every $?X \in \var(R)$. Moreover, given that $P_1$ is a
conjunction of triple patterns and $\var(R) \subseteq \var(P_1)$, we
have that $\mu_1(?X) = \mu(?X)$ for every $?X \in \var(R)$ and, hence,
$\mu_1 \models R$. We deduce that $\mu_1 \in \sem{(P_1 \vc R)}$ and,
hence, $\mu = \mu_1 \cup \mu_2 \in \sem{((P_1 \vc R) \andp
P_2)}$. This concludes the proof of the equivalence of 
$((P_1 \vc R) \andp P_2)$ and $((P_1 \andp P_2) \vc R)$.}

\begin{lemma}\label{lem-and-idempotent}
Let $P$ be a UNION-free graph pattern expression. Then we have that 
\begin{eqnarray*}
(P \andp P) &\equiv& P.
\end{eqnarray*}
\end{lemma}

\sproof{Next we show by induction on the structure of $P$ that for
every RDF database $D$ and pair of mappings $\mu_1, \mu_2 \in \sem{P}$,
if $\mu_1$ and $\mu_2$ are compatible, then $\mu_1 = \mu_2$. It is
easy to see that this condition implies that $(P \andp P) \equiv P$. 

If $P$ is a triple pattern, then the property trivially holds. Assume
first that $P = (P_1 \andp P_2)$, where $P_1$ and $P_2$ satisfy the
condition, that is, if  $\xi, \zeta \in \sem{P_i}$ ($i = 1,2$) and
$\xi$, $\zeta$ are compatible, then $\xi = \zeta$. Let $\mu_1$ and
$\mu_2$ be compatible mappings in $\sem{P}$. Then there exist $\nu_1,
\omega_1 \in \sem{P_1}$ and $\nu_2, \omega_2 \in \sem{P_2}$ such that
$\mu_1 = \nu_1 \cup \omega_1$ and $\mu_2 = \nu_2 \cup \omega_2$. Given
that $\mu_1$ and $\mu_2$ are compatible, we have that $\nu_1$, $\nu_2$
are compatible and $\omega_1$, $\omega_2$ are compatible. Thus, by
induction hypothesis we have that $\nu_1 = \nu_2$ and $\omega_1 =
\omega_2$ and, hence, $\mu_1 = \mu_2$. Second, assume that $P = (P_1
\opt P_2)$, and let $\mu_1$ and $\mu_2$ be compatible mappings in
$\sem{P}$. We consider four cases.  
\begin{enumerate}
\item If there exist $\nu_1,\omega_1 \in \sem{P_1}$ and $\nu_2,
\omega_2 \in \sem{P_2}$ such that $\mu_1 = \nu_1 \cup \omega_1$ and
$\mu_2 = \nu_2 \cup \omega_2$, then we conclude that $\mu_1 = \mu_2$
as in the case $P = (P_1 \andp P_2)$. 

\item If $\mu_1, \mu_2 \in \sem{P_1}$ and both are not compatible
with any mapping in $\sem{P_2}$, then by induction hypothesis we
conclude that $\mu_1 = \mu_2$. 

\item If $\mu_1 \in \sem{P_1}$, $\mu_1$ is not compatible with any
mapping in $\sem{P_2}$, $\mu_2 = \nu_2 \cup \omega_2$, $\nu_2 \in
\sem{P_1}$ and  $\omega_2 \in \sem{P_2}$, then given that $\mu_1$ and
$\mu_2$ are compatible, we have that $\mu_1$ and $\nu_2$ are
compatible. Thus, by induction hypothesis we conclude that $\mu_1 =
\nu_2$ and, therefore, $\mu_1$ is compatible with $\omega_2 \in
\sem{P_2}$, which contradicts our original assumption.

\item If $\mu_1 = \nu_1 \cup \omega_1$, $\nu_1 \in \sem{P_1}$,
$\omega_1 \in \sem{P_2}$,  $\mu_2 \in \sem{P_1}$ and $\mu_2$ is not
compatible with any mapping in $\sem{P_2}$, then we obtain a
contradiction as in the previous case. 
\end{enumerate}
Finally, assume that $P = (P_1 \vc R)$, where $P_1$ satisfy the
condition.  Let $\mu_1$ and $\mu_2$ be compatible mappings in
$\sem{P}$. Then $\mu_1 \in \sem{P_1}$, $\mu_1 \models R$, $\mu_2 \in
\sem{P_2}$, $\mu_2 \models R$ and, thus, $\mu_1 = \mu_2$ by
induction hypothesis. This concludes the proof of the lemma.  
}

\subsection{Proof of Proposition \ref{pro-uni}}
\begin{enumerate}
\item Associative and commutative are consequences of the definitions of operators
$\AND$ and $\UNION$.

\item To prove that $(P_1 \andp (P_2 \uni P_3)) \equiv ((P_1 \andp P_2)
\uni (P_1 \andp P_3))$, we consider two cases. First, we show that for
every RDF database $D$, we have that $\sem{(P_1 \andp (P_2 \uni P_3))}
\subseteq \sem{((P_1 \andp P_2) \uni (P_1 \andp P_3))}$. Assume that
$D$ is an RDF database and that $\mu 
\in \sem{(P_1 \andp (P_2 \uni P_3))}$. Then there exists $\mu_1 \in
\sem{P_1}$ and $\mu_2 \in \sem{(P_2 \uni P_3)}$ such that $\mu_1$ and
$\mu_2$ are compatible and $\mu = \mu_1 \cup \mu_2$. If $\mu_2 \in
\sem{P_2}$, then we have that $\mu = \mu_1 \cup \mu_2 \in
\sem{(P_1 \andp P_2)}$ and, therefore, $\mu \in \sem{((P_1 \andp P_2)
\uni (P_1 \andp P_3))}$. Analogously, if $\mu_2 \in
\sem{P_3}$, then we have that $\mu = \mu_1 \cup \mu_2 \in
\sem{(P_1 \andp P_3)}$ and, therefore, $\mu \in \sem{((P_1 \andp P_2)
\uni (P_1 \andp P_3))}$. Second, we prove that for every RDF database
$D$, we have that $\sem{((P_1 \andp P_2) \uni (P_1 \andp P_3))}
\subseteq \sem{(P_1 \andp (P_2 \uni P_3))}$.   Assume that $D$ is an
RDF database and that $\mu
\in \sem{((P_1 \andp P_2) \uni (P_1 \andp P_3))}$. Then $\mu \in
\sem{(P_1 \andp P_2)}$ or $\mu \in \sem{(P_1 \andp P_3)}$. If $\mu \in
\sem{(P_1 \andp P_2)}$, then we conclude that there exists $\mu_1 \in
\sem{P_1}$ and $\mu_2 \in \sem{P_2}$ such that $\mu_1$ and $\mu_2$ are
compatible and $\mu = \mu_1 \cup \mu_2$. Since $\mu_2 \in \sem{P_2}$,
we have that $\mu_2 \in \sem{(P_2 \uni P_3)}$ and, hence, $\mu = \mu_1
\cup \mu_2 \in \sem{(P_1 \andp (P_2 \uni P_3))}$. If $\mu \in
\sem{(P_1 \andp P_3)}$, then we conclude that there exists $\mu_1 \in
\sem{P_1}$ and $\mu_3 \in \sem{P_3}$ such that $\mu_1$ and $\mu_3$ are
compatible and $\mu = \mu_1 \cup \mu_3$. Since $\mu_3 \in \sem{P_3}$,
we have that $\mu_3 \in \sem{(P_2 \uni P_3)}$ and, therefore, $\mu = \mu_1
\cup \mu_3 \in \sem{(P_1 \andp (P_2 \uni P_3))}$. This concludes the
proof of the equivalence of $(P_1 \andp (P_2 \uni P_3))$ and $((P_1
\andp P_2) \uni (P_1 \andp P_3))$.


\item To prove that $(P_1 \opt (P_2 \uni P_3)) \equiv ((P_1 \opt P_2) \uni (P_1
\opt P_3))$, we consider two cases. First, we show that for every RDF
database $D$, we have that $\sem{(P_1 \opt (P_2
\uni P_3))} \subseteq \sem{((P_1 \opt P_2) \uni (P_1 \opt
P_3))}$. Let $D$ be an RDF database and assume that $\mu \in \sem{(P_1
\opt (P_2 \uni P_3))}$. Then there exists $\mu_1 \in \sem{P_1}$ such that
either (a) there exists $\mu_2 \in \sem{(P_2 \uni P_3)}$ such that
$\mu_1$ and $\mu_2$ are compatible and $\mu = \mu_1 \cup \mu_2$, or
(b) there is no $\mu_2 \in \sem{(P_2 \uni P_3)}$ such that $\mu_1$ and
$\mu_2$ are compatible and $\mu = \mu_1$. In case (a), if $\mu_2 \in
\sem{P_2}$, then $\mu = \mu_1 \cup \mu_2 \in \sem{(P_1 \opt P_2)}$,
and if $\mu_2 \in \sem{P_3}$, then $\mu = \mu_1 \cup \mu_2 \in
\sem{(P_1 \opt P_3)}$. In both cases, we conclude that $\mu \in
\sem{((P_1 \opt P_2) \uni (P_1 \opt P_3))}$. In case (b), we have that
there is no $\mu_2 \in \sem{P_2}$ such that $\mu_1$ and $\mu_2$ are
compatible and, hence, $\mu = \mu_1 \in \sem{(P_1 \opt P_2)}$. We
conclude that $\mu \in \sem{((P_1 \opt P_2) \uni (P_1 \opt P_3))}$. 
Second, we show that for every RDF database $D$, we have that
$\sem{((P_1 \opt P_2) \uni (P_1 \opt P_3))} \subseteq \sem{(P_1 \opt (P_2
\uni P_3))}$. Let $D$ be an RDF database and assume that $\mu \in
\sem{((P_1 \opt P_2) \uni (P_1 \opt P_3))}$.  Then there exists $\mu_1 \in
\sem{P_1}$ such that (a) there exists $\mu_2 \in \sem{P_2}$ such that
$\mu_1$ and $\mu_2$ are compatible and $\mu = \mu_1 \cup \mu_2$, or
(b) there exists $\mu_3 \in \sem{P_3}$ such that $\mu_1$ and $\mu_3$
are compatible and $\mu = \mu_1 \cup \mu_3$, or (c) $\mu = \mu_1$ and
there is neither $\mu_2 \in \sem{P_2}$ compatible with $\mu_1$ nor
$\mu_3 \in \sem{P_3}$ compatible with $\mu_1$. In case (a), given that
$\mu_2 \in \sem{P_2}$, we have that $\mu_2 \in \sem{(P_2 \uni P_3)}$
and, therefore, $\mu = \mu_1 \cup \mu_2 \in \sem{(P_1 \opt (P_2
\uni P_3))}$. In case (b), given that $\mu_3 \in \sem{P_3}$, we have
that $\mu_3 \in \sem{(P_2 \uni P_3)}$ and, therefore, $\mu = \mu_1
\cup \mu_3 \in \sem{(P_1 \opt (P_2 \uni P_3))}$. Finally, in case (c)
we have that there is no $\mu' \in \sem{(P_2 \uni P_3)}$ such that
$\mu_1$ and $\mu'$ are compatible and, therefore, $\mu = \mu_1 \in
\sem{(P_1 \opt (P_2 \uni P_3))}$. This concludes the proof of the
equivalence of $(P_1 \opt (P_2 \uni P_3))$ and $((P_1 \opt P_2) \uni (P_1
\opt P_3))$.

\item To prove that $((P_1 \uni P_2) \opt P_3)) \equiv ((P_1 \opt P_3)
\uni (P_2 \opt P_3))$, we consider two cases. First, we show that for
every RDF database $D$, we have that $\sem{((P_1 \uni P_2) \opt P_3)}
\subseteq \sem{((P_1 \opt P_3) \uni (P_2 \opt P_3))}$. Let $D$ be an
RDF database and assume that $\mu \in \sem{((P_1 \uni P_2) \opt
P_3)}$. Then either (a) there exist $\mu_1 \in \sem{(P_1 \uni P_2)}$
and $\mu_2 \in \sem{P_3}$ such that $\mu_1$ and $\mu_2$ are compatible
and $\mu = \mu_1 \cup \mu_2$, or (b) $\mu \in \sem{(P_1 \uni P_2)}$
and there is no $\mu_3 \in \sem{P_3}$ such that $\mu$ and $\mu_3$ are
compatible. In case (a), if $\mu_1 \in \sem{P_1}$, then $\mu = \mu_1
\cup \mu_2 \in \sem{(P_1 \opt P_3)}$. In case (a), if $\mu_1 \in
\sem{P_2}$, then $\mu = \mu_1 \cup \mu_2 \in \sem{(P_2 \opt P_3)}$. In
case (b), if $\mu \in \sem{P_1}$, then $\mu \in \sem{(P_1 \opt P_3)}$
since $\mu$ is not compatible with any $\mu_3 \in \sem{P_3}$. In case
(b), if $\mu \in \sem{P_2}$, then $\mu \in
\sem{(P_2 \opt P_3)}$ since $\mu$ is not compatible with any $\mu_3
\in \sem{P_3}$. In any of the previous cases, we conclude that $\mu
\in \sem{((P_1 \opt P_3) \uni (P_2 \opt P_3))}$. Second, we show that
for every RDF database $D$, we have that $\sem{((P_1 \opt P_3) \uni
(P_2 \opt P_3))} \subseteq \sem{((P_1 \uni P_2) \opt P_3))}$. Let $D$
be an RDF database and assume that $\mu \in
\sem{((P_1 \opt P_3) \uni (P_2 \opt P_3))}$.  Without loss of
generality, we assume that $\mu \in \sem{(P_1 \opt P_3)}$. Then either
(a) there exists $\mu_1 \in \sem{P_1}$ and $\mu_2 \in \sem{P_3}$ such
that $\mu_1$ and $\mu_2$ are compatible and $\mu = \mu_1 \cup \mu_2$,
or (b) $\mu \in \sem{P_1}$ and there is no $\mu_3 \in \sem{P_3}$ such
that $\mu$ and $\mu_3$ are compatible. In case (a), we have that
$\mu_1 \in \sem{(P_1 \uni P_2)}$ and, hence, $\mu = \mu_1 \cup \mu_2
\in \sem{((P_1 \uni P_2) \opt P_3)}$. In case (b), we have that $\mu
\in \sem{(P_1 \uni P_2)}$ and, therefore, $\mu \in \sem{((P_1 \uni
P_2) \opt P_3)}$ since $\mu$ is not compatible with any $\mu_3 \in
\sem{P_3}$. This concludes the proof of the
equivalence of $((P_1 \uni P_2) \opt P_3)$ and $((P_1 \opt P_3) \uni
(P_2 \opt P_3))$.

\item Clearly, for every RDF database $D$ and built-in condition $R$, we
have that $\{ \mu \in \sem{P_1} \mid \mu \models R\} \subseteq \{ \mu
\in \sem{(P_1 \uni P_2)} \mid \mu \models R\}$ and $\{ \mu \in \sem{P_2}
\mid \mu \models R\} \subseteq \{ \mu \in \sem{(P_1 \uni P_2)} \mid \mu
\models R\}$ since $\sem{P_1} \subseteq \sem{(P_1 \uni P_2)}$ and
$\sem{P_2} \subseteq \sem{(P_1 \uni P_2)}$. Thus, we only need to show
that for every RDF database $D$ and built-in condition $R$, it is the
case that $\sem{((P_1 \uni P_2) \vc R)} \subseteq \sem{((P_1 \vc R)
\uni (P_2 \vc R))}$. Assume that $\mu \in \sem{((P_1 \uni P_2) \vc
R)}$. Then $\mu \in \sem{(P_1 \uni P_2)}$ and $\mu \models R$. Thus,
if $\mu \in \sem{P_1}$, then $\mu \in \sem{(P_1 \vc R)}$, and if $\mu
\in \sem{P_2}$, then $\mu \in \sem{(P_2 \vc R)}$. Therefore, we
conclude that $\mu \in \sem{((P_1 \vc R) \uni (P_2 \vc R))}$. 
\end{enumerate}

\newcommand{\tv}{\text{\tt tv}}
\newcommand{\true}{\text{\tt true}}
\newcommand{\false}{\text{\tt false}}
\newcommand{\ta}{\text{\tt a}}
\newcommand{\tb}{\text{\tt b}}
\newcommand{\tc}{\text{\tt c}}

\subsection{Proof of Theorem \ref{theo-afu}}

It is straightforward to prove that \eval\ is in NP for the case of
graph pattern expressions constructed by using only
$\operatorname{AND}$, $\operatorname{UNION}$ and
$\operatorname{FILTER}$ operators. To prove the NP-hardness of \eval\
for this case, we show how to reduce in polynomial time the
satisfiability problem for propositional formulas in CNF (SAT-CNF) to
our problem. An instance of SAT-CNF is a propositional formula
$\varphi$ of the form:
\begin{eqnarray*}
C_1 \wedge \ldots \wedge C_n,
\end{eqnarray*}
where each $C_i$ ($i \in [1,n]$) is a clause, that is, a disjunction
of propositional variables and negations of propositional
variables. Then the problem is to verify whether there exists a truth
assignment satisfying $\varphi$. It is known that SAT-CNF is
NP-complete \cite{GJ79}.

In the reduction from SAT-CNF, we use a fixed RDF database:
\begin{eqnarray*}
D &=& \{(\ta,\tb, \tc)\}
\end{eqnarray*}
Assume that $x_1$, $\ldots$, $x_m$ is the list of propositional
variables mentioned in $\varphi$. For each $x_i$ ($i \in [1,m]$),
we use SPARQL variables $?X_i$, $?Y_i$ to represent $x_i$ and $\neg
x_i$, respectively. Then for each clause $C$ in $\varphi$ of the form: 
\begin{eqnarray*}
x_{i_1} \vee \cdots x_{i_k} \vee \neg x_{j_1} \vee \cdots \neg
x_{j_\ell},
\end{eqnarray*}
we define a graph pattern $P_C$ as:
\begin{multline*}
((\ta,\tb,?X_{i_1}) \uni \cdots \uni (\ta,\tb,?X_{i_k})\ \uni\\
(\ta,\tb,?Y_{j_1}) \uni \cdots \uni (\ta,\tb,?Y_{j_\ell})),
\end{multline*}
and we define a graph pattern $P_\varphi$ for $\varphi$ as:
\begin{eqnarray*}
(P \andp ((P_{C_1} \andp \cdots \andp P_{C_n}) \vc R)),
\end{eqnarray*}
where:
\begin{eqnarray*}
P &=& ((\ta, \tb, ?X_1) \andp \cdots \andp  (\ta, \tb, ?X_m) \andp\\
&& \phantom{((\ta, \tb, ?X_1) \andp \cdots}(\ta, \tb, ?Y_1) \andp
\cdots \andp (\ta, \tb, ?Y_m)),\\ 
R &=& ((\neg \bound(?X_1) \vee \neg \bound(?Y_1)) \wedge \cdots
\wedge (\neg \bound(?X_m) \vee \neg \bound(?Y_m))).
\end{eqnarray*}
Let $\mu = \{?X_1 \to \tc, \ldots, ?X_m \to \tc, ?Y_1 \to \tc,
\ldots, ?Y_m \to \tc\}$. Then it is straightforward to prove that
$\varphi$ is satisfiable if and only if $\mu \in \sem{P_\varphi}$.

\subsection{Proof of Theorem \ref{theo-cc}}
Membership in PSPACE is a corollary of the membership in PSPACE of the
evaluation problem for first-order logic \cite{V82}. 

To prove the PSPACE-hardness of \eval\ for the case of graph pattern
expressions not containing FILTER conditions, we show how to reduce in
polynomial time the quantified boolean formula problem (QBF) to our
problem. An instance of QBF is a quantified propositional formula
$\varphi$ of the form:
\begin{eqnarray*}
\forall x_1 \exists y_1 \forall x_2 \exists y_2 \forall x_3 \exists
y_3 \cdots \forall x_m \exists y_m\, \psi,
\end{eqnarray*}
where $\psi$ is a quantifier-free formula of the form $C_1 \wedge
\ldots \wedge C_n$, with each $C_i$ ($i \in [1,n]$) being a disjunction
of literals, that is, a disjunction of propositional variables and
negations of propositional variables. Then the problem is to verify
whether $\varphi$ is valid. It is known that QBF is PSPACE-complete
\cite{GJ79}.

In the reduction from QBF, we use a fixed RDF database:
\begin{eqnarray*}
D &=& \{(\ta,\tv,0),\ (\ta,\tv,1),\ (\ta,\false,0),\ (\ta,\true,1)\}.
\end{eqnarray*}
Then for each clause $C$ in $\psi$ of the form
\begin{eqnarray*}
\bigg(\bigvee_{i=1}^k u_i\bigg) \vee \bigg(\bigvee_{j=1}^\ell \neg v_j
\bigg),
\end{eqnarray*}
we define a graph pattern $P_C$ as:
\begin{multline*}
((\ta,\true,?U_1) \uni \cdots \uni (\ta,\true,?U_k)\ \uni\\
(\ta,\false,?V_1) \uni \cdots \uni (\ta,\false,?V_\ell)),
\end{multline*}
and we define a graph pattern $P_\psi$ for $\psi$ as:
\begin{eqnarray*}
(P_{C_1} \andp \cdots \andp P_{C_n}).
\end{eqnarray*}
It is easy to see that $\psi$ is satisfiable if and only if there
exists a mapping $\mu \in \sem{P_\psi}$. In particular, for each
mapping $\mu$, there exists a truth assignment
$\sigma_\mu$ defined as $\sigma_\mu(x) = \mu(?X)$ for every variable
$x$ in $\psi$, such that $\mu \in \sem{P_\psi}$ if and only if
$\sigma_\mu$ satisfies $\psi$.

Now we explain how we represent quantified propositional formula
$\varphi$ as a graph pattern expression $P_\varphi$. We use SPARQL
variables $?X_1$, $\ldots$, $?X_m$ and $?Y_1$, $\ldots$, $?Y_m$ to
represent propositional variables $x_1$, $\ldots$, $x_m$ and $y_1$,
$\ldots$, $y_m$, respectively, and we use SPARQL variables $?A_0$,
$?A_1$, $\ldots$, $?A_m$, $?B_0$, $?B_1$, $\ldots$, $?B_m$ and
operators $\operatorname{OPT}$ and $\operatorname{AND}$ to represent
the quantifier sequence $\forall x_1 \exists y_1 \cdots \forall x_m
\exists y_m$. More precisely, for every $i \in [1,m]$, we define graph
pattern expressions $P_i$ and $Q_i$ as follows:
\begin{align*}
P_i\ :=\ \big(&(\ta, \tv, ?X_1) \andp \cdots \andp (\ta, \tv, ?X_i) \andp\\
&(\ta, \tv, ?Y_1) \andp \cdots \andp (\ta, \tv, ?Y_{i-1}) \andp\\
&(\ta, \false, ?A_{i-1}) \andp (\ta, \true, ?A_i)\big),\\
Q_i\ :=\ \big(&(\ta, \tv, ?X_1) \andp \cdots \andp (\ta, \tv, ?X_i)
\andp\\ 
&(\ta, \tv, ?Y_1) \andp \cdots \andp (\ta, \tv, ?Y_i)
\andp\\
&(\ta, \false, ?B_{i-1}) \andp (\ta, \true, ?B_i)\big),
\end{align*}
and then we define $P_\varphi$ as:
\begin{multline*}
((\ta, \true, ?B_0) \opt (P_1 \opt (Q_1 \opt (P_2 \opt (Q_2 \opt ( \
\cdots\\ (P_m \opt (Q_m \andp P_\psi)) \cdots )))))),
\end{multline*}
Next we show that we can use graph expression $P_\varphi$ to check
whether $\varphi$ is valid. More precisely, we show that
$\varphi$ is valid if and only if $\mu \in \sem{P_\varphi}$, where $\mu$
is a mapping such that $\dom(\mu) = \{?B_0\}$ and $\mu(?B_0) = 1$. 

($\Leftarrow$) Assume that $\mu \in \sem{P_\varphi}$. It is easy to
see that $\sem{P_1} = \{\mu_0, \mu_1\}$, where $\mu_0 = \{?X_1 \to 0,
?A_0 \to 0, ?A_1 \to 1\}$ and $\mu_1 = \{?X_1 \to 1, ?A_0 \to 0, ?A_1
\to 1\}$.  Thus, given that these two mappings are compatible with
$\mu$ and that $\mu \in \sem{P_\varphi}$, there exist mappings $\nu_0$
and $\nu_1$ in $\sem{Q_1}$ such that $\mu_0$, $\nu_0$ are compatible,
$\mu_1$, $\nu_1$ are compatible and
\begin{align}
\tag*{}
\mu_0 \cup \nu_0\ \in\ \sem{(P_1 \opt (Q_1 \opt (&P_2 \opt (Q_2 \opt (
\ \cdots\\
&\label{pspace-eq1}
P_m \opt (Q_m \andp P_\psi)) \cdots )))))},\\
\tag*{}
\mu_1 \cup \nu_1\ \in\ \sem{(P_1 \opt (Q_1 \opt (&P_2 \opt (Q_2 \opt (
\ \cdots\\
\label{pspace-eq2} 
&(P_m \opt (Q_m \andp P_\psi)) \cdots )))))}.
\end{align}
We note that $\nu_0(?X_1) = \mu_0(?X_1) = 0$, $\nu_1(?X_1) =
\mu_1(?X_1) = 0$ and $\nu_0(?Y_1)$, $\nu_1(?Y_1)$ are not necessarily
distinct. 

Since $P_1$ mentions triple $(\ta, \true, ?A_1)$ and $P_2$ mentions
triple $(\ta, \false, ?A_1)$, there is no mapping in $\sem{P_1}$
compatible with some mapping in $\sem{P_2}$. Furthermore, since
$Q_1$ mentions $(\ta, \true, ?B_1)$ and $Q_2$ mentions triple $(\ta,
\false, ?B_1)$, there is no mapping in $\sem{Q_1}$ compatible with
some mapping in $\sem{Q_2}$. Thus, given that (\ref{pspace-eq1})
holds, for every mapping $\zeta \in \sem{P_2}$, we have that if
$\nu_0$ and $\zeta$ are compatible, then there exist $\xi \in
\sem{Q_2}$ such that $\zeta$ and $\xi$ are compatible and
\begin{eqnarray*}
\zeta \cup \xi & \in & \sem{(P_2 \opt (Q_2 \opt ( \ \cdots (P_m \opt
(Q_m \andp P_\psi)) \cdots )))}.
\end{eqnarray*}  
There are two mappings in $\sem{P_2}$ which are compatible with
$\nu_0$:
\begin{eqnarray*}
\mu_{00} &=& \{ ?X_1 \to 0,\  ?X_2 \to 0,\  ?Y_1 \to \nu_0(?Y_1),\  ?A_1 \to
0,\  ?A_2 \to 1\},\\
\mu_{01} &=& \{ ?X_1 \to 0,\ ?X_2 \to 1,\ ?Y_1 \to \nu_0(?Y_1),\ ?A_1 \to
0,\ ?A_2 \to 1\}.
\end{eqnarray*}
Thus, from the previous discussion we conclude that there exist
mappings $\nu_{00}$ and $\nu_{01}$ such that $\mu_{00}$, $\nu_{00}$
are compatible, $\mu_{01}$, $\nu_{01}$ are compatible and
\begin{eqnarray*}
\mu_{00} \cup \nu_{00} &\in& \sem{(P_2 \opt (Q_2 \opt ( \ \cdots (P_m
\opt (Q_m \andp P_\psi)) \cdots )))},\\ 
\mu_{01} \cup \nu_{01} &\in& \sem{(P_2 \opt (Q_2 \opt ( \ \cdots (P_m
\opt (Q_m \andp P_\psi)) \cdots )))}. 
\end{eqnarray*}
Similarly, there are two mapping in $\sem{P_2}$ which are compatible
with $\nu_1$:
\begin{eqnarray*}
\mu_{10} &=& \{ ?X_1 \to 1,\ ?X_2 \to 0,\ ?Y_1 \to \nu_1(?Y_1),\ ?A_1 \to
0,\ ?A_2 \to 1\},\\
\mu_{11} &=& \{ ?X_1 \to 1,\ ?X_2 \to 1,\ ?Y_1 \to \nu_1(?Y_1),\ ?A_1 \to
0,\ ?A_2 \to 1\}.
\end{eqnarray*}
Thus, given that (\ref{pspace-eq2}) holds, we conclude that there exist
mappings $\nu_{10}$ and $\nu_{11}$ such that $\mu_{10}$, $\nu_{10}$
are compatible, $\mu_{11}$, $\nu_{11}$ are compatible and
\begin{eqnarray*}
\mu_{10} \cup \nu_{10} &\in& \sem{(P_2 \opt (Q_2 \opt ( \ \cdots (P_m
\opt (Q_m \andp P_\psi)) \cdots )))},\\ 
\mu_{11} \cup \nu_{11} &\in& \sem{(P_2 \opt (Q_2 \opt ( \ \cdots (P_m
\opt (Q_m \andp P_\psi)) \cdots )))}. 
\end{eqnarray*}
If we continue in this fashion, we conclude that for every $i
\in [2,m-1]$ and $n_1 \cdots n_i \in \{0,1\}^i$, and for the following
mappings in $\sem{P_{i+1}}$:
\begin{align*}
\mu_{n_1 \cdots n_i 0}\ =\ \{ &?X_1 \to n_1,\ \ldots,\ ?X_i \to n_i,\ 
?X_{i+1} \to 0,\\\ 
&?Y_1 \to \nu_{n_1}(?Y_1),\ \ldots,\ ?Y_i \to \nu_{n_1 \cdots
n_i}(?Y_i),\ ?A_{i-1} \to 0,\ ?A_i \to 1\},\\ 
\mu_{n_1 \cdots n_i 1}\ =\ \{ &?X_1 \to n_1,\ \ldots,\ ?X_i \to n_i,\ 
?X_{i+1} \to 1,\\\
&?Y_1 \to \nu_{n_1}(?Y_1),\ \ldots,\ ?Y_i \to
\nu_{n_1 \cdots n_i}(?Y_i),\ ?A_{i-1} \to 0,\ ?A_i \to 1\},
\end{align*}
there exist mappings $\nu_{n_1 \cdots n_i 0}$ and $\nu_{n_1 \cdots n_i
1}$ in $\sem{Q_{i+1}}$ such that $\mu_{n_1 \cdots n_i 0}$, $\nu_{n_1
\cdots n_i 0}$ are compatible, $\mu_{n_1 \cdots n_i 1}$, $\nu_{n_1
\cdots n_i 1}$ are compatible and
\begin{align*}
\mu_{n_1 \cdots n_i 0} \cup \nu_{n_1 \cdots n_i 0}\ \in\ \sem{(P_{i+1}
\opt (&Q_{i+1} \opt ( \ \cdots\\
&(P_m \opt (Q_m \andp P_\psi)) \cdots )))},\\  
\mu_{n_1 \cdots n_i 1} \cup \nu_{n_1 \cdots n_i 1}\ \in\ \sem{(P_{i+1}
\opt (&Q_{i+1} \opt ( \ \cdots\\
&(P_m \opt (Q_m \andp P_\psi)) \cdots )))}.  
\end{align*}
In particular, for every $n_1 \cdots n_m \in \{0,1\}^m$, given that
$\nu_{n_1 \cdots n_m} \in \sem{(Q_m \andp P_\psi)}$, $Q_m$ is a
conjunction of triple patterns and $\var(P_\psi) \subseteq \var(Q_m)$,
we conclude that $\nu_{n_1 \cdots n_m} \in \sem{P_\psi}$. Hence, if
$\sigma_{n_1 \cdots n_m}$ is a truth assignment defined as
$\sigma_{n_1 \cdots n_m}(x) = \nu_{n_1 \cdots n_m}(?X)$ for every
variable $x$ in $\psi$, then $\sigma_{n_1 \cdots n_m}$ satisfies
$\psi$. Thus, given that for every $n_1 \cdots n_m \in \{0,1\}^m$ we
have that:
\begin{center}
\begin{tabular}{lclclcl}
$\mu_{n_1 \cdots n_i}(?X_j)$ &=& $\nu_{n_1 \cdots n_i}(?X_j)$ &=&
$\mu_{n_1 \cdots n_m}(?X_j)$ &\ \ \ \ \ & $i \in [1,m]$ and $j \in [1,i]$,\\ 
$\mu_{n_1 \cdots n_i}(?Y_k)$ &=& $\nu_{n_1 \cdots n_i}(?Y_k)$ &=&
$\mu_{n_1 \cdots n_m}(?Y_k)$ && $ i \in [1,m]$ and $k \in [1,i-1]$,\\
&& $\nu_{n_1 \cdots n_i}(?Y_i)$ &=& $\mu_{n_1 \cdots n_m}(?Y_i)$ && $ i
\in [1,m]$, 
\end{tabular}
\end{center}
we conclude that $\varphi$ is valid.

($\Rightarrow$) The proof that $\varphi$ is valid implies $\mu
\in \sem{P_\varphi}$ is similar to the previous proof.

\subsection{Proof of Theorem~\ref{teo:eval}}
To prove Theorem \ref{teo:eval}, we need some technical lemmas.

\begin{lemma} \label{prop:teoprop}\hfill
\begin{enumerate}
\item\label{prop:joindif}  Let $\Omega_1$, $\Omega_2$, and $\Omega_3$
be set of mappings, then $\Omega_1\mjoin (\Omega_2 \mdif
\Omega_3)\subseteq (\Omega_1 \mjoin \Omega_2)\mdif \Omega_3$. 
  
\item \label{prop:difjoin} Let $\Omega_1$ and $\Omega_2$ be set of
mappings, then $\Omega_1\mdif \Omega_2 = \Omega_1\mdif (\Omega_1\mjoin
\Omega_2)$.
    
\item Let $P_1$, $P_2$ be UNION-free graph pattern expressions and
$\Omega_1$, $\Omega_2$ set of mappings such that $\Omega_1 \subseteq
\sem{P_1}$ and $\Omega_2 \subseteq \sem{P_2}$. Then
$\Omega_1\mpjoin (\Omega_1 \mjoin \Omega_2)=\Omega_1 \mpjoin
\Omega_2$.   
\end{enumerate}
\end{lemma}

\sproof{
\begin{enumerate}
\item Let $\mu\in \Omega_1\mjoin (\Omega_2 \mdif \Omega_3)$ then
$\mu=\mu_1\cup \mu_2$ where $\mu_1\in \Omega_1$, and $\mu_2\in
\Omega_2\mdif \Omega_3$ with $\mu_1$ and $\mu_2$ compatible mappings.
From $\mu_2\in \Omega_2\mdif \Omega_3$ we have that $\mu_2\in
\Omega_2$ and for every mapping $\mu'\in \Omega_3$, $\mu_2$ is
not compatible with $\mu'$.  Note that since $\mu_1$ and $\mu_2$ are
compatible mappings, then $\mu=\mu_1\cup \mu_2\in \Omega_1\mjoin \Omega_2$,
Thus, given that $\mu_2$ is not compatible with any mapping $\mu'\in
\Omega_3$, we conclude that $\mu$ is not compatible with any mapping
$\mu' \in \Omega_3$. Thus, $\mu\in (\Omega_1\mjoin \Omega_2)\mdif
\Omega_3$.  
    
\item First we show that $\Omega_1\mdif \Omega_2 \subseteq
\Omega_1\mdif (\Omega_1\mjoin \Omega_2)$. Let $\mu\in \Omega_1\mdif
\Omega_2$. Then $\mu\in \Omega_1$ and for all 
$\mu'\in \Omega_2$, $\mu$ is not compatible with $\mu'$.  Let $\mu''$ be
any mapping in $\Omega_1\mjoin \Omega_2$, then $\mu''=\mu_1\cup \mu_2$
with $\mu_1\in \Omega_1$, $\mu_2\in \Omega_2$ and then, since $\mu$ is
not compatible with $\mu_2$, necessarily $\mu$ is not compatible with
$\mu''$.  Then $\mu$ is not compatible with every $\mu''\in
\Omega_1\mjoin \Omega_2$, and finally $\mu\in \Omega_1\mdif
(\Omega_1\mjoin \Omega_2)$.  Now we show that $\Omega_1\mdif
(\Omega_1\mjoin \Omega_2) \subseteq \Omega_1\mdif \Omega_2$.  Let
$\mu\in \Omega_1\mdif (\Omega_1\mjoin \Omega_2)$, then $\mu\in
\Omega_1$ and for every $\mu'\in \Omega_1\mjoin \Omega_2$, $\mu$ is
not compatible with $\mu'$.  
Suppose that $\mu$ is compatible with some $\mu''\in \Omega_2$, then
$\mu\cup \mu''\in \Omega_1\mjoin \Omega_2$ and $\mu$ is compatible
with $\mu\cup \mu''$ which is a contradiction with the assumption that
$\mu \in \Omega_1\mdif (\Omega_1\mjoin \Omega_2)$.  Finally, $\mu\in
\Omega_1$ is not compatible with any $\mu''\in \Omega_2$ and then
$\mu\in \Omega_1\mdif \Omega_2$.

\item By definition of $\mpjoin$, we have that 
$\Omega_1 \mpjoin (\Omega_1 \mjoin \Omega_2)=(\Omega_1\mjoin
(\Omega_1\mjoin \Omega_2))\cup (\Omega_1\mdif (\Omega_1\mjoin
\Omega_2))$. By associativity of $\operatorname{AND}$, we have that 
$\Omega_1\mjoin (\Omega_1\mjoin \Omega_2)) = ((\Omega_1 \mjoin
\Omega_1) \mjoin \Omega_2)$, which in turn is equal to $\Omega_1
\mjoin \Omega_2$ since $\Omega_1 \mjoin \Omega_1 = \Omega_1$ by Lemma
\ref{lem-and-idempotent} and the fact that $\Omega_1 \subseteq \sem{P_1}$
and $P_1$ is a UNION-free expression. Furthermore, by property
\ref{prop:difjoin}, we conclude that $(\Omega_1\mdif (\Omega_1\mjoin
\Omega_2)) = \Omega_1 \mdif \Omega_2$ and, therefore, 
$\Omega_1 \mpjoin (\Omega_1 \mjoin \Omega_2)=(\Omega_1\mjoin
(\Omega_1\mjoin \Omega_2))\cup (\Omega_1\mdif (\Omega_1\mjoin
\Omega_2)) = (\Omega_1\mjoin \Omega_2) \cup (\Omega_1\mdif
\Omega_2)=\Omega_1\mpjoin \Omega_2$.
\end{enumerate}}

\begin{lemma} \label{lem:prev}
Let $P$ be a $\UNION$-free graph pattern and $?X\in\var(P)$ a variable of $P$.
If there is a single occurrence of $?X$ that appear in $P$ but in no right hand
size of any $\OPT$ subpattern of $P$, then $?X\in \dom(\mu)$ for all $\mu\in\sem{P}$.
\end{lemma}

\sproof{
First note that the Lemma speaks of \emph{occurrence} of a variable $?X$ and not
of the variable itself.
The intuition of this lemma is that, if an occurrence of $?X$ appear at least
in one of the \emph{mandatory} parts of $P$, then the variable must be bounded in all the mappings
of $\sem{P}$.
The formal proof is by induction in the construction of the pattern.
\begin{enumerate}
\item If $P$ is a triple pattern and $?X\in\var(P)$ then clearly $?X\in\dom(\mu)$ for all
$\mu\in\sem{P}$.
\item Suppose $P=(P_1 \andp P_2)$ .
Then if the occurrence of $?X$ that concern us is in $P_1$ then by induction hypothesis, $?X\in\dom(\mu)$
for all $\mu\in\sem{P_1}$ and then $?X\in\dom(\mu)$ for all $\mu\in\sem{(P_1 \andp P_2)}$.
The case for $P_2$ is the same.
\item Suppose $P=(P_1 \opt P_2)$, then the occurrence of $?X$ that concern us is necessarily in $P_1$.
By induction hypothesis $?X\in\dom(\mu)$ for all $\mu\in\sem{P_1}$ and then by the
definition of $\OPT$, $?X\in\dom(\mu)$ for all $\mu\in\sem{(P_1 \opt P_2)}$.
\end{enumerate}
}

%

\begin{lemma} \label{lem:theproperty}
Let $D$ be an RDF database and $P$ a well-designed graph pattern
expression. Assume that $P'=(P_1\opt P_2)$ is a sub-pattern of $P$ and
$?X$ is a variable such that $?X$ occurs in $P_2$ and $?X$ occurs in
$P$ outside $P'$. Then $?X\in \dom(\mu)$ for every $\mu\in\sem{P_1}$.
\end{lemma}

\sproof{
Let $P'=(P_1 \opt P_2)$ be a subpattern of a well designed graph pattern $P$ such that 
$?X\in \var(P_1)$ and $?X$ occurs outside $P'$.
By the property of $P$ of being well designed, we have that $?X\in\var(P_1)$.
We concetrate now in subpatterns of $P_1$.
Note that because $?X\in \var(P_2)$ and by the hypothesis of $P$ being well designed
for every occurrence of $?X$ in the right hand size of an $\OPT$ subpattern
of $P_1$ there is an occurrence of $?X$ in the left hand size of the same $\OPT$ subpattern.
The last statement imply that there is necessarily an occurrence of $?X$ that is not at the right hand
size of any of the $\OPT$ subpatterns of $P_1$, because if it were not the case $P_1$
would have an infinite number of occurrence of $?X$ 
(we would never stop applying the property of well designed pattern).
Then applying Lemma~\ref{lem:prev} we obtain that for every $\mu\in\sem{P_1}$, $?X\in\dom(\mu)$,
completing the proof.
}

\begin{lemma} \label{lem:superproperty}
Let $D$ be an RDF database and $P$ a well-designed graph pattern
expression. Suppose that $P'$ is a sub-pattern of $P$ and $?X$ is a
variable such that $?X$ occurs in $P'$ and $?X$ occurs in $P$ outside
$P'$. Then $?X\in \dom(\mu)$ for every $\mu\in\sem{P'}$.
\end{lemma}

\sproof{By induction on $P'$.
\begin{enumerate}
\item If $P'$ is a triple pattern $t$, then $?X\in \dom(\mu)$ for every $\mu\in\sem{t}$.

\item Let $P'=(P_1 \andp P_2)$. If $?X\in \var(P_1)$, then
by induction hypothesis $?X\in \dom(\mu)$ for every $\mu\in\sem{P_1}$,
and then $?X\in \dom(\nu)$ for every $\nu\in\sem{(P_1 \andp P_2)}$.
If $?X\in \var(P_2)$ the proof is similar.

\item Let $P'=(P_1 \opt P_2)$. If $?X\in \var(P_1)$ then
by induction hypothesis $?X\in \dom(\mu)$ for every 
$\mu\in\sem{P_1}$, and then $?X\in \dom(\nu)$ for every
$\nu\in\sem{(P_1 \opt P_2)}$.  If $?X\in \var(P_2)$, then given that
$P$ is a well-designed graph pattern expression and $?X$ occurs in $P$
outside $P'$, we have that $?X \in \var(P_1)$. We conclude that $?X\in
\dom(\nu)$ for every $\nu\in\sem{(P_1 \opt P_2)}$ as in the previous
case.

\item Let $P'=(P_1 \vc R)$. Then $?X\in \var(P_1)$ and, thus, by induction
hypothesis $?X\in \dom(\mu)$ for every $\mu\in\sem{P_1}$.  Now by
definition $\sem{(P_1 \vc R)}\subseteq \sem{P_1}$ and, therefore,
$?X\in \dom(\nu)$ for every $\nu\in\sem{(P_1 \vc R)}$
\end{enumerate}}

\aproof{Theorem \ref{teo:eval}}
{
We will prove that during the execution of $Eval_D(\,\cdot\,)$, for every
call $Eval_D(P,\Omega)$ it holds that $Eval_D(P,\Omega)=\Omega\mjoin \sem{P}$.
This immediatelty implies that $Eval_D(P)=\sem{P}$ because $Eval_D(P)=Eval_D(P,\{\mu_\emptyset\})$.

The property trivially holds when
$\Omega=\emptyset$ since $Eval_D(P,\Omega) = \emptyset =
\emptyset\mjoin \sem{P}$. Thus, we assume that $\Omega\not=\emptyset$.
Now the proof goes by induction on $P$.
\begin{itemize}
\item If $P$ is a triple pattern $t$, then $Eval_D(P, \Omega) = \Omega
\mjoin \sem{t}$. 

\item Suppose that $P=(P_1 \andp P_2)$. Computing $Eval_D(P,\Omega)$
is equivalent to compute $Eval_D({P_2},Eval_D({P_1},\Omega))$ then by
induction hypothesis, $Eval_D(P, \Omega) = Eval_D({P_2}, \Omega \mjoin
\sem{P_1})=\Omega \mjoin \sem{P_1} \mjoin \sem{P_2}=\Omega\mjoin
\sem{(P_1 \andp P_2)}$.
  
\item Suppose that $P=(P_1 \opt P_2)$.  Computing $Eval_D(P,\Omega)$
is equivalent to compute $Eval_D({P_1},\Omega) \mpjoin
Eval_D({P_2},Eval_D({P_1},\Omega))$ and then by induction hypothesis
$Eval_D(P,\Omega)=(\Omega \mjoin \sem{P_1}) \mpjoin (\Omega \mjoin
\sem{P_1} \mjoin \sem{P_2})$. Thus, we need to show that
\begin{eqnarray*}
(\Omega \mjoin \sem{P_1}) \mpjoin (\Omega \mjoin \sem{P_1} \mjoin
\sem{P_2})=\Omega \mjoin (\sem{P_1} \mpjoin \sem{P_2}).
\end{eqnarray*}
First we show that $\Omega \mjoin (\sem{P_1} \mpjoin \sem{P_2})
\subseteq (\Omega \mjoin \sem{P_1}) \mpjoin (\Omega \mjoin \sem{P_1}
\mjoin \sem{P_2})$.  Let $\mu\in \Omega \mjoin (\sem{P_1} \mpjoin
\sem{P_2})$ then $\mu=\mu_1\cup\mu_2$ where $\mu_1\in \Omega$,
$\mu_2\in (\sem{P_1} \mpjoin \sem{P_2})$, and $\mu_1$, $\mu_2$ are
compatible mappings.  We consider two cases:
\begin{enumerate}
\item[(a)] $\mu_2\in \sem{P_1}\mjoin \sem{P_2}$. Then $\mu\in \Omega
\mjoin (\sem{P_1} \mjoin \sem{P_2})$ and, hence, by commutativity and
associativity of the $\operatorname{AND}$ operator and Lemma
\ref{lem-and-idempotent}, we have that $\mu \in (\Omega \mjoin \sem{P_1})
\mjoin (\Omega \mjoin \sem{P_1} \mjoin \sem{P_2})\subseteq (\Omega
\mjoin \sem{P_1}) \mpjoin (\Omega \mjoin \sem{P_1} \mjoin \sem{P_2})$. 
   
\item[(b)] $\mu_2\in \sem{P_1}\mdif \sem{P_2}$. Then $\mu \in \Omega \mjoin
(\sem{P_1}\mdif \sem{P_2})\subseteq (\Omega \mjoin \sem{P_1})  \mdif
\sem{P_2}$ (by Lemma \ref{prop:teoprop} \ref{prop:joindif}) and, thus,
$\mu \in (\Omega\mjoin \sem{P_1})\mdif(\Omega\mjoin
\sem{P_1}\mjoin \sem{P_2})$ (by Lemma \ref{prop:teoprop}
\ref{prop:difjoin} and conmutativity and associativity of the
$\operatorname{AND}$ operator). We conclude that $\mu\in (\Omega
\mjoin \sem{P_1}) \mpjoin (\Omega \mjoin \sem{P_1} \mjoin \sem{P_2})$.
\end{enumerate}
   
Now we show that $(\Omega \mjoin \sem{P_1}) \mpjoin (\Omega \mjoin
\sem{P_1} \mjoin \sem{P_2}) \subseteq \Omega \mjoin (\sem{P_1} \mpjoin
\sem{P_2})$.  By the definition of $\mpjoin$\!\!\!, it is sufficient to show
that $(\Omega \mjoin \sem{P_1}) \mjoin (\Omega \mjoin \sem{P_1} \mjoin
\sem{P_2}) \subseteq \Omega \mjoin (\sem{P_1} \mpjoin \sem{P_2})$, and
that $(\Omega \mjoin \sem{P_1}) \mdif (\Omega \mjoin \sem{P_1} \mjoin
\sem{P_2}) \subseteq \Omega \mjoin (\sem{P_1} \mpjoin \sem{P_2})$:
\begin{enumerate}
\item[(a)] By commutativity and associativity of the
$\operatorname{AND}$ operator and Lemma \ref{lem-and-idempotent}, we
have that $(\Omega \mjoin \sem{P_1}) \mjoin (\Omega \mjoin \sem{P_1}
\mjoin \sem{P_2})=\Omega \mjoin \sem{P_1} \mjoin \sem{P_2}\subseteq
\Omega\mjoin (\sem{P_1}\mpjoin \sem{P_2})$. 
    
\item[(b)] By Lemma \ref{prop:teoprop} \ref{prop:difjoin}, to show
that $(\Omega \mjoin \sem{P_1}) \mdif (\Omega \mjoin \sem{P_1} \mjoin
\sem{P_2}) \subseteq \Omega \mjoin (\sem{P_1} \mpjoin \sem{P_2})$
is equivalent to show that $(\Omega \mjoin \sem{P_1}) \mdif \sem{P_2}
\subseteq \Omega \mjoin (\sem{P_1}\mpjoin \sem{P_2})$.  Let $\mu\in
(\Omega \mjoin \sem{P_1})$ be such that for every $\mu'\in
\sem{P_2}$, $\mu$ is not compatible with $\mu'$.  Then $\mu=\mu_1\cup
\mu_2$ with $\mu_1\in \Omega$, $\mu_2\in \sem{P_1}$, and $\mu_1$,
$\mu_2$ compatible mappings. Furthermore, for every $\mu'\in \sem{P_2}$,
$\mu_1\cup \mu_2$ is not compatible with $\mu'$.  Suppose that $\mu_2$
is not compatible with any $\mu' \in \sem{P_2}$, then $\mu_2\in
\sem{P_1}\mdif\sem{P_2}\subseteq \sem{P_1}\mpjoin \sem{P_2}$, and then
$\mu=\mu_1\cup \mu_2\in \Omega\mjoin (\sem{P_1}\mpjoin \sem{P_2})$.
Suppose now that $\mu_2$ is compatible with some $\nu \in \sem{P_2}$,
but $\mu_1$ is not compatible with $\nu$.  Then there exists a
variable $?X\in \dom(\mu_1)$ such that $?X\in \dom(\nu)$ and
$\mu_1(?X)\not=\nu(?X)$. Since $\mu_2$ is compatible with both $\mu_1$
and $\nu$, we have that $?X \not\in \dom(\mu_2)$.  This implies that
$?X$ is in the domain of a mapping in $\Omega$ since $\mu_1 \in
\Omega$ and, hence, $?X$ is defined outside $P=(P_1 \opt
P_2)$. Furthermore, $?X\in \var(P_2)$ since $?X \in \dom(\nu)$ and
there exists a mapping $\omega = \mu_2 \in \sem{P_1}$ such that $?X
\not\in \dom(\omega)$, which contradicts Lemma
\ref{lem:theproperty}. This conclude the proof of the inclusion
$(\Omega \mjoin \sem{P_1}) \mdif (\Omega \mjoin \sem{P_1} \mjoin
\sem{P_2}) \subseteq \Omega \mjoin (\sem{P_1} \mpjoin \sem{P_2})$. 
\end{enumerate} 
  
\item Suppose that $P=(P_1 \vc R)$. Computing $Eval_D(P,\Omega)$
results in the set of mappings $\{\mu\in Eval_D(P_1,\Omega)\;|\;
\mu\models R\}$.  By induction hypothesis this set is equal to
$\{\mu\in \Omega \mjoin \sem{P_1}\;|\; \mu\models R\}$. Thus, we need
to show that this set is equal to $\Omega \mjoin \sem{(P_1 \vc R)}$. 
First, assume that $\nu \in \Omega \mjoin \sem{(P_1 \vc R)}$. Then $\nu
= \nu_1 \cup \nu_2$ with $\nu_1 \in \Omega$, $\nu_2 \in \sem{(P_1 \vc
R)}$ and $\nu_1$, $\nu_2$ compatible mappings. Since $\nu_2 \in
\sem{(P_1 \vc R)}$ we have that $\nu_2 \in \sem{P_1}$ and $\nu_2
\models R$. Next we show that $\nu \models R$. By contradiction,
assume that $\nu \not\models R$. Then given that $\nu_2 \models R$ and
$\nu = \nu_1 \cup \nu_2$, there is a variables $?X \in \var(R)$ such
that $?X \in \dom(\nu)$ but $?X \not\in \dom(\nu_2)$. But this implies
that $?X \in \dom(\nu_1)$ and, therefore, $?X$ occurs outside $P$
since $\nu_1 \in \Omega$. We conclude that $?X$ occurs in $P$, $?X$
occurs outside $P$ and there exists a mapping $\omega  = \nu_2 \in
\sem{P}$ such that $?X \not\in \dom(\omega)$, which contradicts Lemma
\ref{lem:superproperty}. Thus, we conclude that $\nu \models R$ and,
therefore, $\nu = \nu_1 \cup \nu_2 \in \{\mu\in \Omega \mjoin
\sem{P_1}\;|\; \mu\models R\}$.  Second, assume that $\nu
\in \{\mu\in \Omega \mjoin \sem{P_1}\;|\; \mu\models R\}$. Then $\nu
\models R$ and $\nu = \nu_1 \cup \nu_2$ with $\nu_1 \in \Omega$,
$\nu_2 \in \sem{P_1}$ and $\nu_1$, $\nu_2$ compatible mappings. Next
we show that $\nu_2 \models R$. By contradiction, assume that $\nu_2
\not\models R$. Then given that $\nu \models R$ and $\nu = \nu_1 \cup
\nu_2$, we have that there exists variable $?X \in \var(R)$ such that
$?X \in \dom(\nu)$ but $?X \not\in \dom(\nu_2)$. But this implies
that $?X \in \dom(\nu_1)$ and, therefore, $?X$ occurs outside $P_1$
since $\nu_1 \in \Omega$. We conclude that $?X$ occurs in $P_1$ since
$\var(R) \subseteq \var(P_1)$, $?X$ occurs outside $P_1$ and there
exists a mapping $\omega  = \nu_2 \in \sem{P_1}$ such that $?X \not\in
\dom(\omega)$, which contradicts Lemma \ref{lem:superproperty}. Thus,
we conclude that $\nu_2 \models R$ and, 
therefore, $\nu_2 \in \sem{(P_1 \vc R)}$. Hence, we deduce that $\nu =
\nu_1 \cup \nu_2 \in \Omega \mjoin \sem{(P_1 \vc R)}$. This concludes
the proof of the theorem.

\end{itemize}
}


\subsection{Proof of Proposition~\ref{prop:andopt}}

First we show that for every subpattern $(P_1 \andp (P_2 \opt P_3))$ of a well designed pattern $P$, 
it holds that $(P_1 \andp (P_2 \opt P_3)) \equiv ((P_1 \andp P_2) \opt P_3).$

\sproof{
To simplify the notation we will suppose that $\mu_1\in\sem{P_1}$, $\mu_2\in\sem{P_2}$, and $\mu_3\in\sem{P_3}$.
\begin{itemize}
\item First $\sem{(P_1 \andp (P_2 \opt P_3))}\subseteq \sem{((P_1 \andp P_2) \opt P_3)}$.
Let $\mu\in \sem{(P_1 \andp (P_2 \opt P_3))}=\sem{P_1}\mjoin (\sem{P_2}\mpjoin \sem{P_3})$.
Then $\mu=\mu_1\cup\mu'$ with $\mu_1$ and $\mu'$ compatible mappings, and 
$\mu'\in\sem{P_2}\mpjoin \sem{P_3}$, depending on $\mu'$ there are two cases:
  \begin{itemize}
  \item If $\mu'\in \sem{P_2}\mjoin \sem{P_3}$ then $\mu\in \sem{P_1}\mjoin (\sem{P_2}\mjoin \sem{P_3})$,
  and then $\mu\in (\sem{P_1}\mjoin \sem{P_2})\mjoin \sem{P_3}\subseteq\sem{((P_1 \andp P_2) \opt P_3)}$
  \item If $\mu'\in \sem{P_2}\mdif \sem{P_3}$ then $\mu'\in\sem{P_2}$ and is incompatible with every 
  $\mu_3\in\sem{P_3}$, then $\mu=\mu_1\cup \mu'$ is incompatible with $\mu_3$ and then 
  $\mu\in (\sem{P_1}\mjoin \sem{P_2}) \mdif \sem{P_3}\subseteq (\sem{P_1}\mjoin \sem{P_2}) \mpjoin \sem{P_3}$
  and then $\mu \in \sem{((P_1 \andp P_2) \opt P_3)}$.
  \end{itemize}
\item Now $\sem{((P_1 \andp P_2) \opt P_3)} \subseteq \sem{(P_1 \andp (P_2 \opt P_3))}$.
Let $u\in\sem{(P_1 \andp P_2) \opt P_3))}=(\sem{P_1}\mjoin\sem{P_2})\mpjoin \sem{P_3}$.
There are two cases:
  \begin{itemize}
  \item $\mu\in(\sem{P_1}\mjoin\sem{P_2})\mjoin \sem{P_3}=(\sem{P_1}\mjoin\sem{P_3})\mjoin\sem{P_2}$
  then $\mu\in\sem{((P_1 \andp P_2) \opt P_3)}$.
  \item $\mu\in(\sem{P_1}\mjoin\sem{P_2})\mdif \sem{P_3}$, then $\mu=\mu_1\cup\mu_2$ with 
  $\mu_1$ and $\mu_2$ compatible mappings and for every $\mu_3$, $\mu_1\cup\mu_2$ is incompatible with $\mu_3$.
  Suppose first that $\mu_2$ is incompatible with $\mu_3$, then 
  $\mu_2\in \sem{P_2}\mdif \sem{P_3}\subseteq \sem{P_2}\mpjoin\sem{P_3}$ and then
  $\mu_1\cup\mu_2\in \sem{P_1}\mjoin (\sem{P_2}\mpjoin\sem{P_3})=\sem{(P_1 \andp (P_2 \opt P_3))}$.
  Suppose now that $\mu_1$ is incompatible with $\mu_3$, then there exists a variable $?X\in\dom(\mu_1)$,
  $?X\in\dom(\mu_3)$ such that $\mu_1(?X)\not=\mu_3(?X)$.
  This last statement imply that $?X\in\var(P_1)\cap\var(P_3)$ and then because $P$ is well designed
  by Lemma~\ref{lem:theproperty} we obtain $?X\in\dom(\mu_2)$ and because $\mu_2$ is compatible with $\mu_1$
  we have that $\mu_2(?X)\not=\mu_3(?X)$.
  Finally $\mu_2\in \sem{P_2}\mdif \sem{P_3}\subseteq \sem{P_2}\mpjoin \sem{P_3}$, and then
  $\mu=\mu_1\cup\mu_2 \in \sem{(P_1 \mjoin (P_2 \mpjoin P_3))}$.
  \end{itemize}
\end{itemize}
}

Now we show that for every subpattern $((P_1 \opt P_2) \opt P_3)$ of a well designed pattern $P$,
it holds that $((P_1 \opt P_2) \opt P_3) \equiv ((P_1 \opt P_3) \opt P_2).$

\sproof{
\begin{itemize}
\item First $\sem{((P \opt P_1) \opt P_2)}\subseteq \sem{((P \opt P_2) \opt P_1)}$.
Let $\mu\in \sem{((P \opt P_1) \opt P_2)}$ then $\mu\in (\sem{P}\mpjoin \sem{P_1})\mpjoin\sem{P_2}$.
Suppose that $\mu\in (\sem{P} \mpjoin \sem{P_1})\mjoin \sem{P_2}$, there are two cases:
\begin{itemize}
  \item $\mu \in (\sem{P}\mjoin \sem{P_1}) \mjoin \sem{P_2}\subseteq (\sem{P}\mjoin \sem{P_2}) \mjoin \sem{P_1}\subseteq \sem{((P \opt P_2) \opt P_1)}$.
  
  \item $\mu \in (\sem{P}\mdif \sem{P_1}) \mjoin \sem{P_2} \subseteq (\sem{P} \mjoin \sem{P_2}) \mdif \sem{P_1}$, by proposition~\ref{prop:teoprop}~\ref{prop:joindif}, then  $ \mu \in \sem{((P \opt P_2) \opt P_1)}$.
\end{itemize}
Suppose now that $\mu\in (\sem{P} \mpjoin \sem{P_1})\mdif \sem{P_2}$
There are two cases: (i assume $\mu'\in\sem{P}, \mu_1\in\sem{P_1}, \mu_2\in\sem{P_2})$.
  \begin{itemize}
  \item $\mu\in (\sem{P} \mjoin \sem{P_1}) \mdif \sem{P_2}$, then $\mu=\mu' \cup \mu_1$ compatibles mappings, and for every $\mu_2$, $\mu'\cup \mu_1$ is incompatible with $\mu_2$.
  If $\mu'$ is incompatible with $\mu_2$ then $\mu'\in\sem{P}\mdif\sem{P_2}$ and then $\mu'\cup\mu_1 \in (\sem{P}\mdif \sem{P_2})\mjoin \sem{P_1})$ and then $\mu \in \sem{((P \opt P_2) \opt P_1)}$.
  Suppose that $\mu_1$ is incompatible with $\mu_2$, then there is $?X$ such that $\mu_1(?X)\not=\mu_2(?X)$.
  Then $?X\in\var(P_1)\cap\var(P_2)$ and because the whole pattern is well designed, 
  by Lemma~\ref{lem:theproperty} we obtain that $?X\in\mu'$ and by $\mu'$ compatible with $\mu_1$ we obtain that $\mu'(?X)\not=\mu_2(?X)$, and then $\mu'$ is incompatible with $\mu_2$.
  Then $\mu\in\sem{((P \opt P_2) \opt P_1)}$.
  \item $\mu\in (\sem{P} \mdif \sem{P_1}) \mdif \sem{P_2}$, then $\mu\in \sem{P}$ and is such that for all $\mu_1$ and for all $\mu_2$, $\mu$ is incompatible with $\mu_1$ and $\mu_2$, and then $\mu\in (\sem{P} \mdif \sem{P_2}) \mdif \sem{P_1}\subseteq \sem{((P \opt P_2) \opt P_1)}$.
\end{itemize}
\item Now we show that $\sem{((P \opt P_2) \opt P_1)}\subseteq \sem{((P \opt P_1) \opt P_2)}$.
Let $\mu\in \sem{((P \opt P_1) \opt P_2)}$ then $\mu\in (\sem{P}\mpjoin \sem{P_2})\mpjoin\sem{P_1}$.
(again i assume $\mu'\in\sem{P}, \mu_1\in\sem{P_1}, \mu_2\in\sem{P_2})$.
Suppose that $\mu\in (\sem{P}\mpjoin \sem{P_2})\mjoin\sem{P_1}$, there are two cases:
\begin{itemize}
  \item $\mu \in (\sem{P}\mjoin \sem{P_2})\mjoin\sem{P_1} \subseteq \sem{((P \opt P_1) \opt P_2)}$.
  \item $\mu \in (\sem{P}\mdif \sem{P_2}) \mjoin\sem{P_1} \subseteq (\sem{P}\mjoin \sem{P_1})\mdif \sem{P_2}$ by prop.~\ref{prop:teoprop}~\ref{prop:joindif}
  and then $\mu \in \sem{((P \opt P_1) \opt P_2)}$.
\end{itemize}
Suppose now that $\mu\in (\sem{P}\mpjoin \sem{P_2})\mdif\sem{P_1}$, there are two cases:
\begin{itemize}
  \item $\mu\in (\sem{P}\mjoin \sem{P_2}) \mdif \sem{P_1}$, then $\mu=\mu'\cup \mu_2$ compatible mappings such that for every $\mu_1\in \sem{P_1}$, $\mu'\cup\mu_2$ is incompatible with $\mu_1$.
  If $\mu'$ is incompatible with $\mu_1$ then $\mu'\in\sem{P}\mdif\sem{P_1}$ and then $\mu'\cup \mu_2\in (\sem{P}\mdif\sem{P_1})\mjoin \sem{P_2}\subseteq (\sem{P}\mdif\sem{P_1})\mpjoin \sem{P_2}$ and then $\mu \in \sem{((P \opt P_1) \opt P_2)}$.
  If $\mu_2$ is incompatible with $\mu_1$ then there exists a variable $?X\in\dom(\mu_1)\cap\dom(\mu_2)$ such that $\mu_1(?X)\not=\mu_2(?X)$.
  Then $?X\in\var(P_1)\cap\var(P_2)$ and because the whole pattern is well designed, 
  by Lemma~\ref{lem:theproperty} we obtain that $?X\in\mu'$ and by $\mu'$ compatible with $\mu_2$ we obtain that $\mu'(?X)\not=\mu_1(?X)$, and then $\mu'$ is incompatible with $\mu_1$.
  Then $\mu\in\sem{((P \opt P_1) \opt P_2)}$.
  
  \item $\mu\in (\sem{P}\mdif \sem{P_2}) \mdif \sem{P_1}\subseteq (\sem{P}\mjoin \sem{P_1}) \mdif \sem{P_2}\subseteq \sem{((P \opt P_1) \opt P_2)}$
\end{itemize}
\end{itemize}
}

To finish the proof we must show that replacing the respective 
equivalences do not affect the property of $P$ of being well
designed.
Let $(P_1 \andp (P_2 \opt P_3))$ be a subpattern of $P$.
Well designed says that, if a variable $?X$ occurs outside 
$(P_2 \opt P_3)$ and inside $P_3$ then it occurs in $P_2$.
Suppose that this is the case and that $?X$ occurs outside 
$(P_1 \andp (P_2 \opt P_3))$, then because $?X$ occurs in 
$?P_2$ then $?X$ occurs in $(P_1 \andp P_2)$ and the pattern 
$P'$ obtained from $P$ by replacing $(P_1 \andp (P_2 \opt P_3))$
by $(P_1 \andp P_2) \opt P_3))$ is well designed.
Suppose now that $?X$ occurs in $P_1$ but does not occur outside 
$(P_1 \andp (P_2 \opt P_3))$, then $?X$ does not occur outside 
$((P_1 \andp P_2) \opt P_3)$ and then the pattern obtained from $P$ is well designed.

The proof for $P'=((P_1 \opt P_2) \opt P_3)$ is similar.
There are various cases for variables occurring inside $P_2$, $P_3$.
\begin{itemize}
\item $?X$ occurs in $P_2$ and in $P_3$, 
\item $?X$ occurs in $P_2$ and outside $P'$ but not in $P_3$, 
\item $?X$ occurs in $P_3$ and outside $P'$ but not in $P_2$,
\end{itemize}
in all cases because $P$ is well designed $?X$ occurs in $P_1$
and then the pattern obtained from $P$ replacing $P'$ by 
$((P_1 \opt P_3) \opt P_2)$ is well designed.


\subsection{Proof of Theorem~\ref{teo:norm}}
To prove Theorem~\ref{teo:norm} we use the following Lemma.
In the Lemma we use rewriting concepts and results (see~\cite{Baader}).

\begin{lemma}\label{teo:E}
Let us consider the  theory $E$ formed by the equations of 
associativity and commutativity for $\AND$ (Proposition~\ref{pro-uni}), and equation 
\[
((X \opt Y) \opt Z) \equiv ((X \opt Z) \opt Y)
\]
Then the rule
\begin{eqnarray}
  (X \andp (Y \opt Z)) \longrightarrow ((X \andp Y) \opt Z)  \label{ruler}
\end{eqnarray}
is $E$-terminating and $E$-confluent in the set of well designed patterns, 
and hence has $E$-normal forms in the set of well designed patterns. 
\end{lemma}

\sproof{
\begin{enumerate}
\item First we prove that rule (\ref{ruler}) is terminating.
Consider the measure 
\[
m(P): \text{ number of }\OPT\text{ inside }\AND\text{-trees in the parsing of }P.
\]
Then clearly the theory $E$ keeps $m(P)$ constant.
Let $P'$ and $P''$ be the left and right hand side in rule~(\ref{ruler}) respectively,
then $m(P') > m(P'')$.
Hence successive application of rule~(\ref{ruler}) must terminate.
 
\item Now we prove that rule~(\ref{ruler}) is $E$-locally confluent.
Note that the only critical pair (see~\cite{Baader}) is: 
$((P_1 \opt P_2) \andp (P_3 \opt P_4))$ 
Then it only left to check that both applications of rule~(\ref{ruler})
\[
      (((P_1 \opt P_2) \andp P_3) \opt P_4)
\]
and 
\[
      (((P_3 \opt P_4) \andp P_1) \opt P_2)
\]
can be rewritten to a common term using the axioms of $E$ and the rule~(\ref{ruler}):
\[
\begin{array}{rcl}
(((P_1 \opt P_2) \andp P_3) \opt P_4) & \stackrel{E}{\equiv} & ((P_3 \andp (P_1 \opt P_2)) \opt P_4) \\
& \stackrel{(\ref{ruler})}{\rightarrow} & (((P_3 \andp P_1) \opt P_2) \opt P_4) \\
& \stackrel{E}{\equiv} & (((P_1 \andp P_3) \opt P_2) \opt P_4)
\end{array}
\]

\[
\begin{array}{rcl}
(((P_3 \opt P_4) \andp P_1) \opt P_2) & \stackrel{E}{\equiv} & ((P_1 \andp (P_3 \opt P_4)) \opt P_2) \\
& \stackrel{(\ref{ruler})}{\rightarrow} & (((P_1 \andp P_3) \opt P_4) \opt P_2) \\
& \stackrel{E}{\equiv} & (((P_1 \andp P_3) \opt P_2) \opt P_4)
\end{array}
\]
\end{enumerate}
}

Theorem~\ref{teo:norm} follows from the existence of $E$ normal forms for rule~(\ref{ruler}),
 and the application of~(\ref{ruler}) and $E$ identities to well designed graph patterns.

\end{document}